\documentclass[12pt]{article}
\usepackage{setspace}
\usepackage{amsmath}
\usepackage{amsfonts}
\usepackage{amssymb}
\usepackage{amsthm}
\usepackage{graphicx}
\usepackage{booktabs}
\usepackage{float}
\usepackage{subfig}
\usepackage[countmax]{subfloat}
\usepackage[round]{natbib}
\usepackage{url}

\newtheorem{claim}{{\bf \sc Claim}}
\newtheorem{theorem}{{\bf \sc Theorem}}

\newtheorem{proposition}{{\bf \sc Proposition}}
\newtheorem{observation}{{\bf \sc Observation}}

\newtheorem{remark}{{\bf \sc Remark}}

\def\eproof{\hbox{\hskip3pt\vrule width4pt height8pt depth1.5pt}}
\begin{document}

\title{Networks of Military Alliances, Wars, and International Trade }
\author{Matthew O. Jackson and Stephen M. Nei\thanks{%
Department of
Economics, Stanford University, Stanford, California 94305-6072 USA.  Jackson is also
an external faculty member at the Santa Fe Institute and a member of CIFAR.
Emails: jacksonm@stanford.edu and snei@stanford.edu.  We thank Antonio Cabrales,
Matt Elliott, Jim Fearon, Ben Golub, Rachel Kranton, John Ledyard, and Massimo Morelli, as well as various seminar participants, for
helpful comments.  We gratefully acknowledge
financial support from the NSF under grants SES-0961481 and SES-1155302 and
from grant FA9550-12-1-0411 from the AFOSR and DARPA, and ARO MURI award No.
W911NF-12-1-0509.}}
\date{Draft date: June 2015}
\maketitle

\begin{abstract}

We investigate the role of networks of alliances in preventing (multilateral) interstate wars.  We first show that, in the absence of international trade, no network of alliances is peaceful and stable.  We then show that international trade induces peaceful and stable networks:  trade increases the density of alliances so that countries are less vulnerable to attack and also reduces countries' incentives to attack an ally.  We present historical data on wars and trade, noting
that the dramatic drop in interstate wars since
1950, and accompanying densification and stabilization of alliances, are consistent with the model.
Based on the model we also examine some specific relationships, finding that countries that have high levels of trade with their allies are less likely to be involved in wars with any other countries (including allies and non-allies), and that increased trade between two countries decreases the chance that they end up in a war.


\end{abstract}

\thispagestyle{empty}
\setcounter{page}{0}

\pagebreak

\section{Introduction}
\label{sec:intro}

\begin{quote}
Wars are caused by undefended wealth.
\emph{ Ernest Hemingway (repeated by Douglas MacArthur in lobbying to fortify the Philippines in the 1930's\footnote{See the biography by Bob Considine, source for Chapter 1: Deseret News, Feb 24, 1942.})}
\end{quote}

\begin{quote}There is only one thing worse than fighting with allies, and that is fighting without them.  \emph{Winston Churchill, April 1, 1945}\footnote{Arthur Bryant, Triumph in the West, 1943-1946 (London: Grafton Books, 1986), 349}\end{quote}

The enormous costs of war make it imperative to understand the conditions under which wars are likely to occur, and the ways in which they can be prevented.  Although much is known about bilateral conflicts, there is no formal theory of how networks of multilateral international relationships foster and deter interstate wars.  In this paper we introduce a model of networks of military alliances and international
trade that can serve as a foundation for study of international alliance structure and conflict.

In terms of background, the history of the networks of international alliances is rich and nuanced.
Arranging multiple alliances to ensure world peace found perhaps it most famous proponent in Otto von Bismarck and his belief that the European states could be allied in ways that would maintain a peaceful balance of power.\footnote{
E.g., see \citet*{taylor69}. 
}
The alliances that emerged were briefly stable following the unification and expansion of Germany that took place up through the early 1870s, but were ultimately unable to prevent World War I.  Indeed, many world conflicts involve multiple countries allied together in defensive and offensive groups, from the shifting alliances of the Peloponnesian and Corinthian wars of ancient Greece to the Axis and Allies of World War II, and so studying the fabric of alliances is
 necessary for understanding international (in)stability.
 Based on the ``Correlates of War'' data set, between 1823 and 2003, 40 percent of wars with more than 1000 casualties involved more than two countries, and indeed some of the most destructive (e.g., the World Wars,  Korean War, Vietnam,...) involved multilateral conflicts.\footnote{This is based on
 the COW data 
 for which there is data regarding initiators of the war,
 which we then couple with other data for our analysis.  This does not even include the Napoleonic wars, as the data begin afterwards.  Also, there are some wars that might
 be thought of as civil, but that involve substantial interstate conflict: e.g., the Second Congo War, the Russian Revolution, etc.}
 Most importantly, this is really a network problem.  As we detail in Section \ref{CC}, multilateral wars never involved cliques (fully allied coalitions of more
  than two countries) against cliques.  Out of the 95 wars between 1823 and 2003 that qualify as having at least one side with three or more countries, {\sl none} of them involved a clique versus a clique.  Thus, a network approach of understanding alliances, rather than a coalitional one (in which countries are partitioned into allied groups) is warranted.  As we also show, a network approach meshes well with patterns of international trade, which are far from coalitional and instead involve rich network patterns.

\begin{figure}[!h]
	\centering
	\includegraphics[angle=270, scale=0.35]{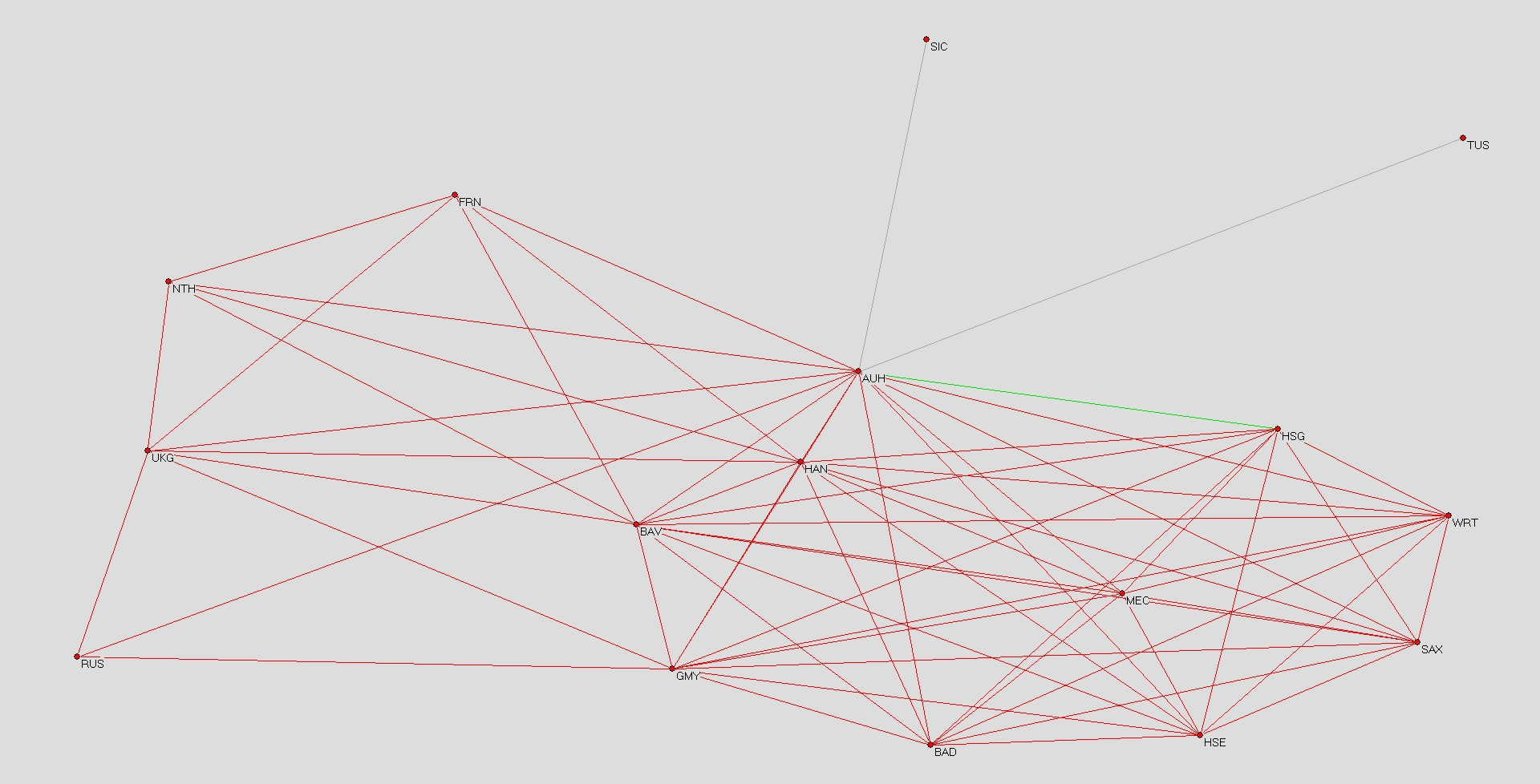}
	\caption{Network of Alliances, 1815, red for multilateral alliance, grey for bilateral alliance, green for both}
	\label{figure:1815}
\end{figure}

\begin{figure}[!h]
	\centering
	\includegraphics[angle=270, scale=0.35]{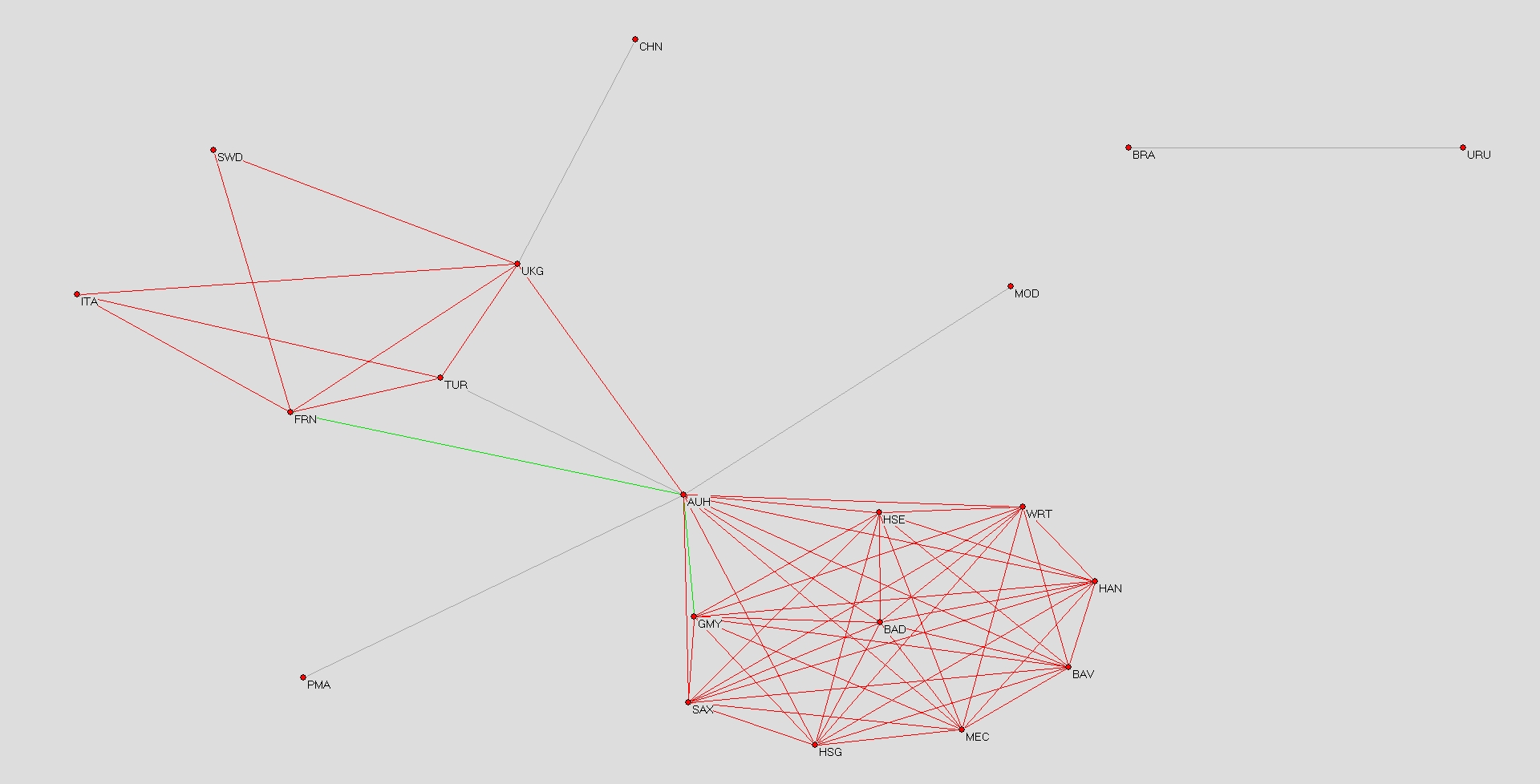}
	\caption{Network of Alliances, 1855, red for multilateral alliance, grey for bilateral alliance, green for both}
	\label{figure:1855}
\end{figure}

\begin{figure}[!h]
	\centering
	\includegraphics[angle=270, scale=0.35]{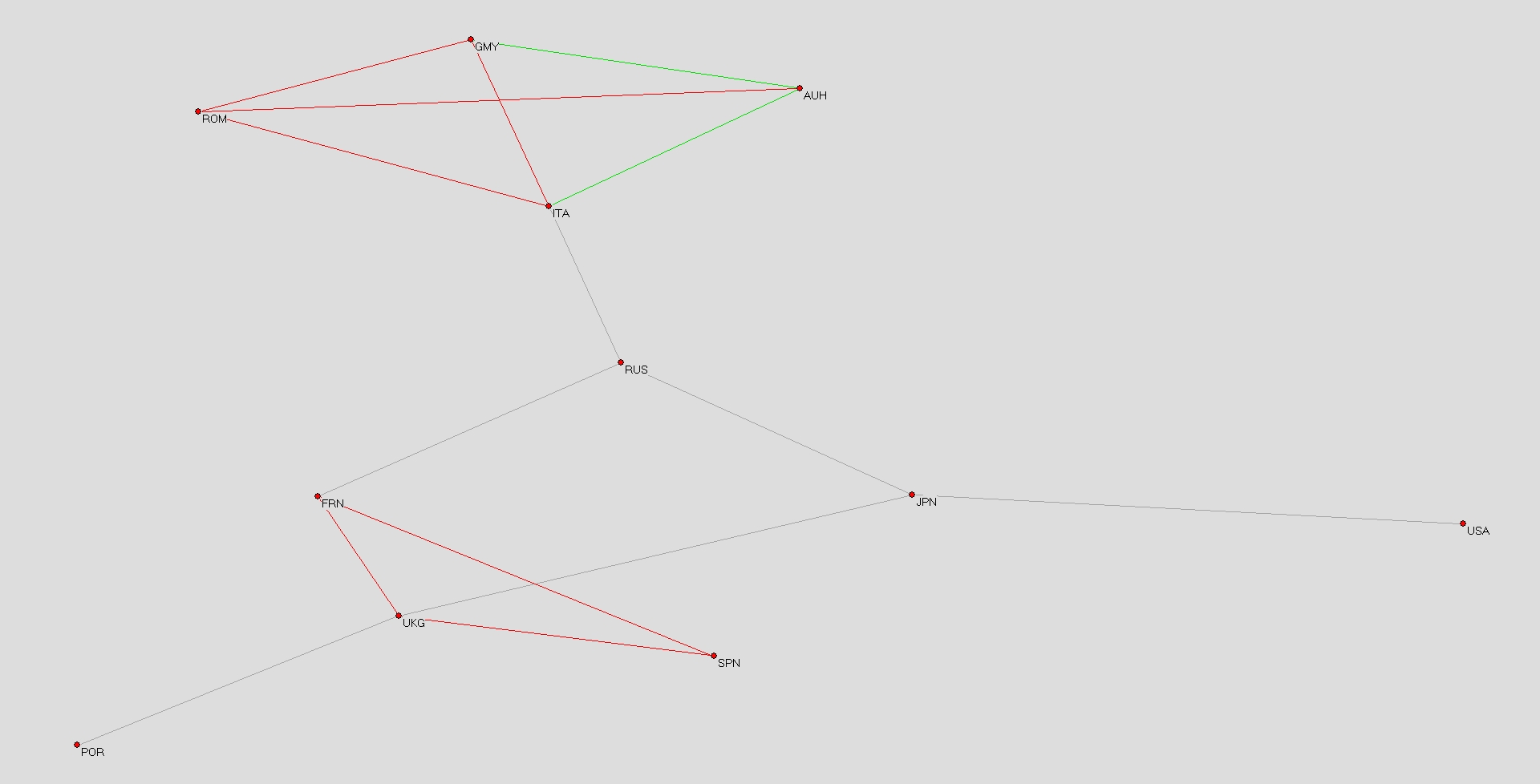}
	\caption{Network of Alliances, 1910, red for multilateral alliance, grey for bilateral alliance, green for both}
	\label{figure:1910}
\end{figure}

\begin{figure}[!h]
	\centering
	\includegraphics[angle=270, scale=0.35]{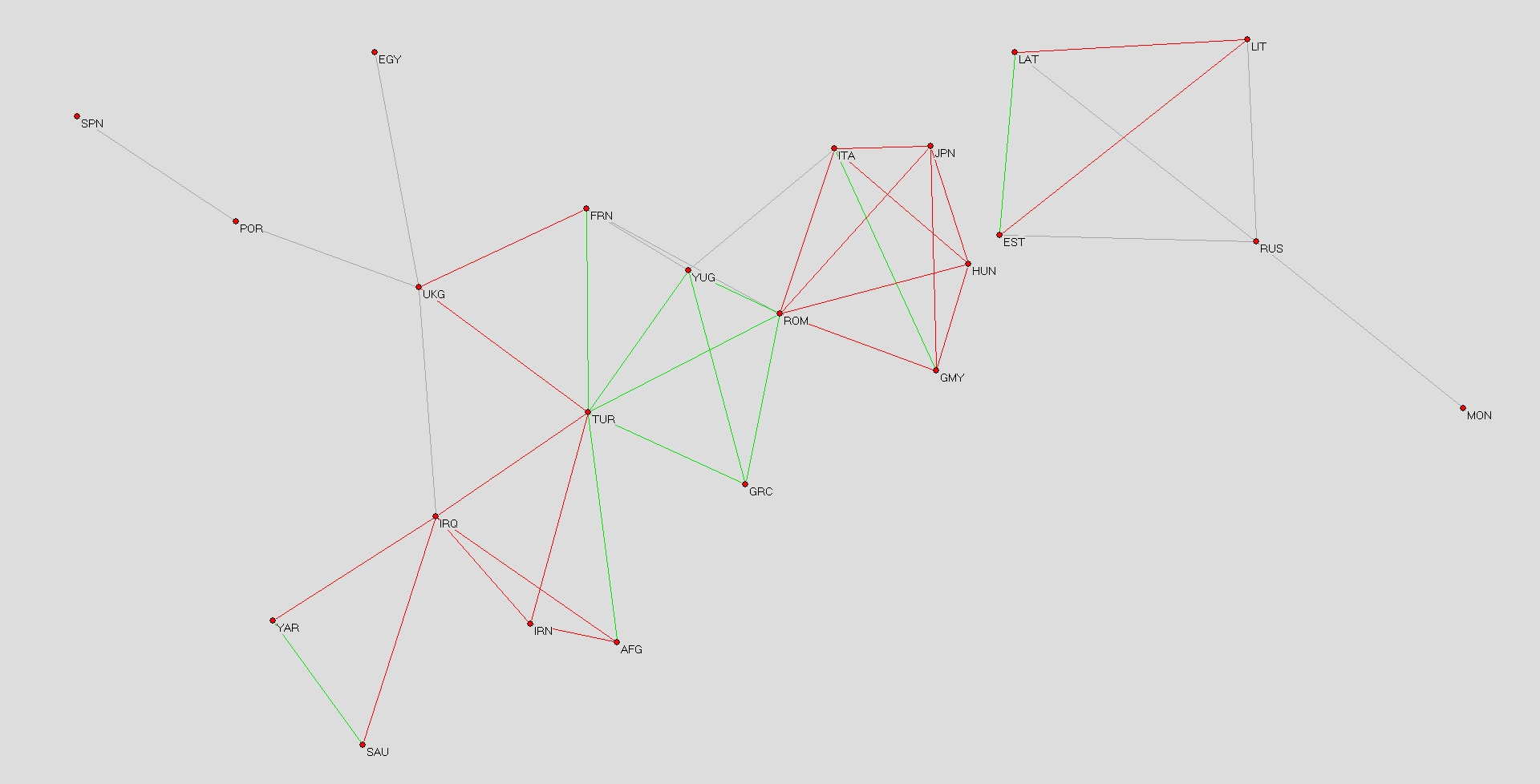}
	\caption{Network of Alliances, 1940, red for multilateral alliance, grey for bilateral alliance, green for both}
	\label{figure:1940}
\end{figure}

\begin{figure}[!h]
	\centering
	\includegraphics[angle=270, scale=0.35]{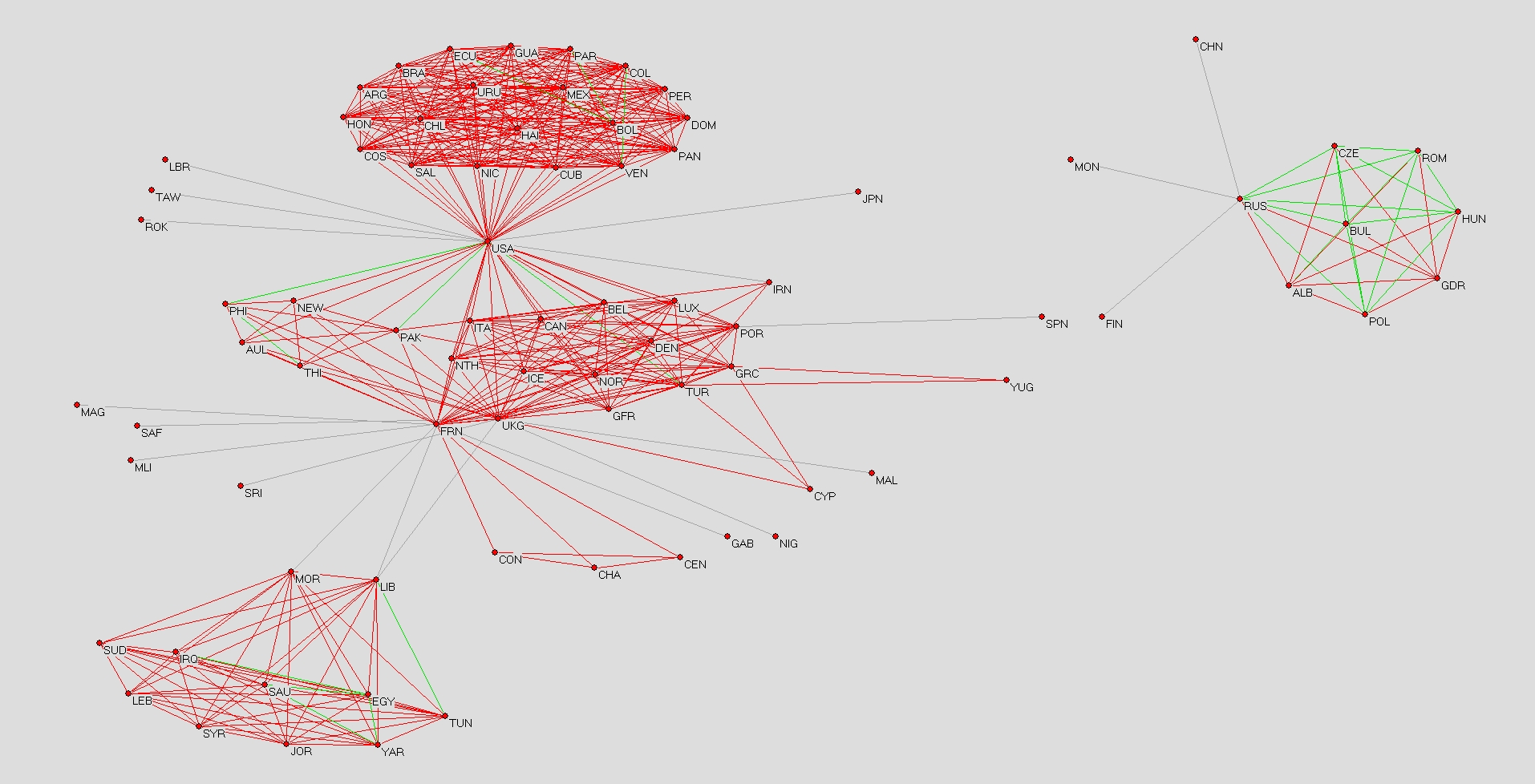}
	\caption{Network of Alliances, 1960, red for multilateral alliance, grey for bilateral alliance, green for both}
	\label{figure:1960}
\end{figure}

The historical background on the networks of alliances between the early 1800s and the present basically breaks into two periods, with the break occurring after World War II\footnote{We use alliance data reported by the Alliance Treaty Obligation and Provisions Project \url{atop.rice.edu}, including alliances marked as containing at least one of a defensive, offensive, or consultation provision.}.  This can be seen in Figures \ref{figure:1815} through \ref{figure:2000}.  The early period (pre-1950) involved relatively sparse, very dynamic and unstable networks, and many wars. The time series of these early networks exhibits rapid shifts, with very different alliances existing decade to decade.  The later period (post-1950) involves increasingly dense, highly stable networks, and relatively very few wars. The networks stabilize and become substantially denser and with alliances that are separated by continent and ideology - there are large cliques, corresponding to large geographical areas, which are bridged by a few larger states.  As a preview of the analysis in Section \ref{empirics}, between 1816 and 1950 a country had on average
2.525 alliances, while from
1951 to 2003 this grows by a factor of more than four to 10.523.
In terms of turnover: between 1816 to 1950, for an alliance that is present in one year, there is only a $0.695$ probability that it will still be present five years later.
In contrast, during the period of  1950 to 2000 the frequency increases to $0.949$.

To gain insights into networks of alliances and the incidence of wars, we model the incentives of countries to attack each other, to form alliances, and to trade with each other.
We first present a base concept of networks that are stable against wars from a purely military point of view, when trade is ignored.   A group of countries can attack some other country if all members of the attacking coalition share a mutual ally.  The idea is that alliances represent the necessary means for coordinating military action.    A country that is attacked can be defended by its allies.   A country is {\sl vulnerable} if there is some aggressor country and a coalition of its allies whose collective military strength outweighs that of the country and its remaining allies who are not in the attacking coalition (adjusted by a parameter that captures technological considerations that may give an advantage to offensive or defensive forces).\footnote{
We also explore other definitions based on other rules of which
connections are needed between countries in order to attack and/or defend,
and show that the results hold for those alternative definitions. }

In addition to not having any vulnerable countries, endogenizing the networks is essential to understanding stability.
Thus, we define a concept of {\sl war-stable} networks that accounts for the incentives of countries to form and drop alliances.  We build upon the concept of pairwise stability of \citet*{jacksonw1996}, adapting it to the current setting.

\noindent In particular, a network of alliances is {\sl war-stable} if three conditions are met:
 \begin{itemize}
 \item first, no country is vulnerable to a successful attack by others,
 \item second, no two countries that are not allied could form an alliance that would allow them, together with allies, to successfully attack another country, and
 \item third, each existing alliance serves a purpose - any country that deletes any of its alliances would end up being vulnerable.
 \end{itemize}
This concept embodies the simple principles that countries prefer to win a war and not to lose one, and that alliances are costly and so should
serve some purpose in order to be maintained.

It turns out that there are no war-stable networks, even with this definition that imposes minimal requirements. The tension is
 understood as follows. Requiring that
countries not be vulnerable to attack and having every alliance serve some purpose leads networks to be relatively sparse - with each country having a few alliances but a network that is not overly dense.  However, this can make a country
susceptible to some of its allies joining forces and defeating it.  Essentially, the pressure to economize on alliances conflicts with stability against the formation of new alliances, which leads to instability and would suggest chaotic dynamics.

\begin{figure}
	\centering
	\includegraphics[height=3.0in]{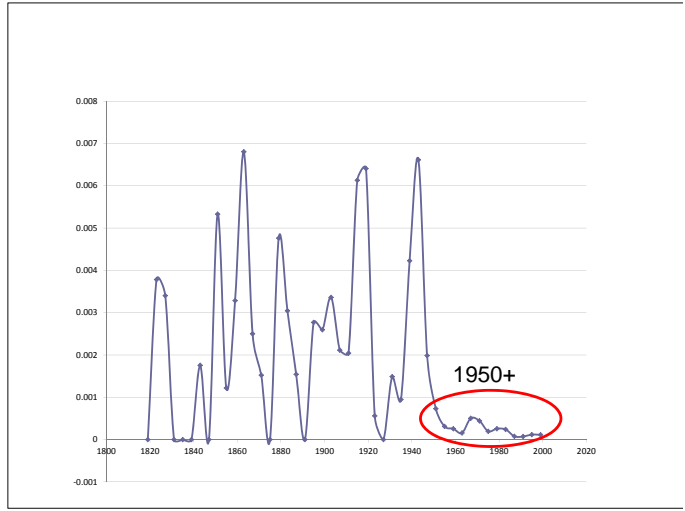}
	\caption{Wars per pair of countries by year, 1820-2000. (Participant level observations from COW MIDB 4.01 dataset, number of entries with hostility level 5 divided by number of pairs of countries in COW State System Membership)}
	\label{figure:warsperdyad}
\end{figure}

This instability provides insights into the constantly shifting structures and recurring wars that occurred throughout the nineteenth and first half of the twentieth centuries.\footnote{There was a relatively quiet period prior to World War I that was prosperous and
during which trade increased and there was some temporary stability.  However, stability was partly due to the relative asymmetries
in the strengths of Germany and Austria-Hungary compared to France and Russia (and the fact that Germany had already gained much territory from those countries); but this subsided as France and Russia regained
the relative strength that they had lost during the nineteenth century.  Interlocking trade was not yet sufficient to prevent the `Great War' from occurring, and the alliance structure proved far from stable.}
Wars, however, have greatly subsided in parallel with the huge increase of trade (which was partly driven by the introduction of containerized shipping in the 1960s, which greatly decreased costs):   Between 1820 and 1959 each pair of countries averaged .00056 wars per year, while from 1960 to 2000 the average was .00005 wars per year, less than {\sl one tenth} as much.
We see this pattern quite clearly in Figure \ref{figure:warsperdyad}.\footnote{Even if one measures this per country rather than per pair of potential combatants, the decrease has been more than threefold, as discussed in Section \ref{warempirics}.}
These changes also follow the advent of nuclear weapons, which impacted the technology of war.  However, we show that nuclear weapons cannot lead to
stability in the absence of trade, as our model allows for quite arbitrary asymmetries between the military capabilities of countries and offensive or defensive advantages in wars - and instability ensues for {\sl any} specification of technology and relative military capabilities.   Indeed, in order to capture the actual patterns that have emerged one must add other considerations - such as trade considerations - since the base model shows that  all networks of alliances are unstable
with nuclear weapons but without trade.\footnote{The cold war was accompanied by a (temporary) change to a form of bilateralism, that we come back to in Section \ref{sec:concl}.  Again, to understand the accompanying peace and patterns of alliances, international trade is instrumental.}

Thus, the second part of our analysis is to enrich the base model to include international trade.
Indeed, there has been a rapid increase in global trade since World War II (partly coincident with the growth of container shipping among other stimuli).  The empirical relationship between war and trade is an active area of research, with strong suggestions (e.g., \citet*{martin08}) that  network concerns may be important.
So, we introduce a concept of a network of alliances being {\sl war and trade stable},  which allows countries to form alliances for either economic or military considerations.  In this richer model, an alliance allows countries to trade with each other and to coordinate military activities, and so can be formed for either reason.
This restores existence of networks of alliances that are stable against the addition or deletion of alliances.
Trade provides two helpful incentives:  first it provides economic motivations to maintain alliances, and the resulting denser network of alliances then has a deterrent effect; and second, it can reduce the incentives of a country to attack another since trade will be disrupted.  This reduces the potential set of conflicts and, together with the denser networks, allows for a rich family of stable networks that can exhibit structures similar to networks we see currently.

We provide some results on the existence and structure of war and trade stable networks of alliances, showing that structures similar to those observed over the past few decades are economically stable under apparently reasonable parameters.
It is important to note that another dramatic change during the post-war period was the introduction of nuclear weapons, which changes the technology of war and is generally thought to have greatly increased the defensive advantage to those with such weapons.\footnote{Another change has been in the number of democracies.  The endogeneity of such changes makes
it difficult to factor in, but even accounting for democratization, trade still seems to be an important factor, as discussed by \citet{oneal1999}.}  Our model suggests that although world-wide adoption of nuclear weapons could stabilize things in the absence of trade, it would result in an empty network of alliances as the stable network.  To explain the much denser and more stable networks in the modern age along with the paucity of war in a world where nuclear weapons are limited to a small percentage of countries, our model points to the enormous growth in trade as a big part of the answer.

We close the paper with some discussion of this role that the growth in trade has played in reducing wars over the past half century, and how this relates to the advent of the nuclear age.
The model provides some specific predictions for some of the mechanisms that decrease wars: including trade with allies (making it more likely that allies will aid a country and less likely that they will be part of an attacking coalition against that country), and that increasing trade between any two countries lowers their gains from war with each other making it less likely that they will be at war at any time.
We show that both of these effects are observed in the data.

Before proceeding, let us say a few words about how this paper contributes to the study of war.
The literature on war provides many rationales for why wars occur.
Our analysis here fits firmly into what has become a ``rationalist'' tradition based on cost and benefit analyses by rational actors, with roots seen in writings such as \citet*{hobbes} Leviathan, and has become
the foundation for much of the recent international relations literature.\footnote{Background can be found in \cite{fearon95} and \citet*{jackson11}.}

To our knowledge, there are no previous models of conflict that
game-theoretically model networks of alliances between multiple
agents/countries based on costs and benefits of wars.
\footnote{There is a literature
that adapts the balance theory of \citet{heider46} to
examine network patterns of enmity
(e.g., \citet{hiller12,rietzker13,koenig14}).
The ideas in those works build upon notions of the
form that ``the enemy of my enemy is my friend,'' and
are quite different from the sort of cost-benefit analysis underlying the military and trading alliances considered here.}
There are previous models of coalitions in conflict settings (e.g., see \citet{bloch2012} for a survey).
Here, network structures add several things to the picture.  Our model is very much in a similar rationalist perspective of the literature that examines group conflict (e.g., \citet{esteban1999,esteban2001,esteban2003}), but enriching it to admit network structures of alliances and of international trade.
This allows us to admit patterns that are consistent with the networks of alliances that are actually observed, which are far from being partitions (e.g., the U.S. is currently allied with both Israel and Saudi Arabia, Pakistan and India, just to mention a couple of many prominent examples).   More importantly, our Theorem \ref{econstable} provides a first model
in which such non-partitional such structures are stable and provide insight into peace.  Moreover, as we already mentioned above, the observed patterns of wars and of alliances are not partitional, and so this provides an important advance in moving the models towards matching observed patterns of wars, trade and alliances.
Our model thus serves as a foundation upon which one can eventually build more elaborate analyses of multilateral interstate alliances, trade, and wars.
It is also important to emphasize that the network of international trade is complex and can in fact be
stable (and prevent conflict) precisely because it cuts {\sl across}
coalitions.  This is in contrast to coalitional models that generally
predict only the grand coalition can be stable or that very exact balances
are possible (e.g., see \citet*{blochss2006}).
Again, this is something illustrated
in our Theorem \ref{econstable}, and which does not exist in the previous literature.  Finally, our model illuminates the relationships between international trade, stable network structures, and peace, something not appearing in the previous literature - as the previous literature that involves international trade and conflict generally revolves around bilateral reasoning or focuses on instability and armament (e.g., \citet*{garfinkelss2014}) and does not address the questions that we address here.

The complex relationship between trade and conflict is the subject of a growing empirical literature (e.g., \citet*{barbieri96,mansfield97,martin08,glick10,hegreor2010}).  The literature not only has
to face challenges of endogeneity and causation, but also of substantial heterogeneity in relationships, as well as geography, and the level of conflict.  The various correlations between conflict and trade are complex, and a model such as ours that combines military and economic incentives, and others that may follow, can provide structure with which to interpret some of the empirical observations, which we discuss in the following section.

\section{Empirics of Trade and Wars}\label{empirics}

We begin by presenting some empirical background that motivates the development of our model.

\subsubsection{Trends in Military Alliance Networks}

Marked differences occur between the military alliances we see in the ATOP data over time.\footnote{
The number of countries in the data set grows over time, and so everything we do adjusts on a per country
basis, as otherwise the trends are even magnified further.  The number of states in 1816 was 23, in
in 1950 it was 75, and by
2003 it reached 192.}
There are two major changes that we see in the period before and after the Second World War.  These changes are also easy to see in the Figures in Section \ref{snapshots}.

The first major change is that there is a great deal of turnover in alliances, which constantly shift in
the period from 1816 to 1950.
In particular, let us do a simple calculation: how frequently do alliances disappear?  Specifically, consider an alliance that is present in year $t$, and calculate the frequency with which it is also present
in year $t+5$.  Doing this for each year from 1816 to 1950, we find the frequency to be $0.695$.
When doing this for each year from 1950 to 2003 the frequency becomes $0.949$.   Thus, there is an almost one-third chance that any given alliance disappears in the next five years in the pre-WWII period, and then only a five-percent chance that any given alliance at any given time will disappear within the next five years in the post-WWII period.

The second major change is that the network of alliances greatly densifies.
Between 1816 and 1950 a country had on average
2.525 alliances  (standard deviation 3.809).  If one drops the WWII decade of the 1940s during which most countries were allied in one of two blocks, then this number drops down even further to
1.722 between 1816 to 1940 (standard deviation of 1.366).
During the period of 1951 to 2003 this grows by a factor of more than four to 10.523 (standard deviation of 1.918).   Thus, there are substantially more alliances per country in the post war than the pre-war period.

To summarize, countries have just a couple of alliances on average and those alliances rapidly turned over in the pre-WWII period; while in contrast countries
form on average more than ten alliances and do not turn them over in the post-WWII period.

\subsubsection{Trends of Wars and Conflicts}
\label{warempirics}

Another trend that is quite evident is that the number of wars per country has decreased dramatically post World War II, and that this decrease comes even though the number of countries has increased - so that there are many more pairs of countries that could go to war.
For example, the average number of wars per pair of countries per year from 1820 to 1959 was .00056  while from 1960 to 2000 it was .00005, less than a tenth of what it was in the previous period.  We saw this in Figure \ref{figure:warsperdyad}.

This finding is robust to when the cut takes place:  from 1820 to 1949 it was .00059  while from 1950 to 2000 it was .00006,  from 1820 to 1969 it was .00053  while from 1970 to 2000 it was .00005.
If one looks at wars per country instead of per pair of countries, then from 1820 to 1959 it was .012  while from 1960 to 2000 it was .004.   One could also include all Militarized Interstate Disputes (MID2-5) instead of just wars (MID5s - involving at least 1000 deaths). In that case, from 1820 to 1959 there are .006 MIDs per pair of countries while from 1960 to 2000 there were .003.
Thus, the decrease in wars is quite robust to the way in which this is measured.

It is also interesting to note that with the exceptions of the Korean and Vietnam wars, which had major cold-war considerations, (as well as the anomalous Falklands war),  the 24 other MID 5's since 1950 generally involved lesser-developed (lower-trade) countries as the major protagonist on at least one (and often both) sides of the dispute.  Moreover, major trading partners at the time do not appear on opposite sides of the dispute.

\subsubsection{Trade}

International trade has had two major periods of growth over the last two centuries, one in the latter part of the nineteenth century and beginning of the twentieth, disrupted by the first world war, and then picking up again after the second world war, recovering to its 1914 levels through the 1960s and then continuing to grow at an increasing rate thereafter.   In particular, \citet*{estevadeordal03} finds that
trade per capita grew by more than 1/3 in each
decade from 1881 to 1913, while it grew only 3 percent
per decade from 1913 to 1937.  Table \ref{table-trade}, from \citet*{krugman95},\footnote{The figure for 2012 is directly from the
World Bank indicator (http://data.worldbank.org/topic/private-sector?display=graph, December 11, 2013), from which \citet*{krugman95} quotes the other numbers.}
provides a view of this dynamic.\footnote{\citet*{dean04} provide an overview of changes in the level of world trade in relation to world output over the course of the 20th century, while \citet*{estevadeordal03} looks at the period 1870 to 1939.}

\begin{table}[h]
\caption{World merchandise exports as percent of GDP: \citet*{krugman95}}
\label{table-trade}
\begin{center}
\begin{tabular}{|c||c|c|c|c|c|c|c|c|}
\hline
Year &  1850 &1880 &1913 &1950 &1973 &1985 &1993 & 2012\\
\hline
Percent &5.l &9.8 &11.9 &7.1 &11.7 &14.5 &17.1 & 25.3\\
\hline
\end{tabular}
\end{center}
\end{table}

The trade has been further bolstered or accompanied by the advent of container shipping, and other advances in tanker and shipping technology, which dramatically decreased costs of trade, as well as increases in world per capita income.
\citet*{hummels07} looks at the interaction between transportation costs and international trade, while \citet*{bernhofen13} and \citet*{rua12} investigate the rise of containerization and its spread through international shipping.  The relative correlations between income and trade and transportation costs and trade have been open to some debate.   \citet*{baier01}, looking at trade between OECD countries from the late 1950s through the late 1980s, argues that decreasing transportation costs explains 8 percent of the growth in trade, with the lion's-share of the increase (67 percent) correlating with increased incomes.
Regardless of the source, trade has increased dramatically over time, and especially post World War II, where it has increased by almost a factor of four.

\subsubsection{Relations Between Trade and Wars and Confounding Variables}

Putting these two trends together, we see that the decrease in wars is mirrored by an increase in trade.  The percentage of trade varies mainly between 5 and 12 percent from 1850 to 1959 and between 12 and 25 percent from 1960 onwards.

There are many papers that have investigated the empirical relationship between conflict and trade at a more dyadic level, and as one might expect causation and the specifics of the relationships are difficult to disentangle.
Indeed, \Citet*{barbieri96} -- investigating the period 1870 to 1938 in Europe and
 including conflicts that fall substantially short of war -- finds that although low to moderate levels of economic interdependence may be accompanied by a decrease in military conflicts; high levels of economic interdependence can be accompanied by increased incidence of conflicts.
This inversion is nuanced, as \citet*{martin08} -- looking at trade and militarized disputes over the period 1950-2000 -- find that an increase in bilateral trade between two countries correlates with a decreased likelihood of these countries entering military dispute with each other, while an increase in one of the country's multilateral trade (i.e. an overall increase in a country's trade share without an increase in the bilateral trade between the two countries) leads to an increased likelihood of war between the pair.  The definition of dispute is broader than that of war and could include posturing for bargaining purposes, or simply the increase in contact that accompanies trade leading to an increase in minor incidents.
\citet*{oneal1999} provide evidence that with a careful examination of proximity and trade, trade significantly reduces conflict, although again such results are correlations and not proof of causation.\footnote{See also \citet*{hegreor2010}, who provide some evidence for the aspect of our model that costs of conflict deter wars.}

These numbers do not imply causation as there are many confounding variables in the relationship between trade and wars. So, although there was an unprecedented growth in trade post World War II, coincident with an unprecedented drop in the frequency of wars, there were also many technological changes, as well as an increase in income and wealth levels world-wide, and a growth in the number of democracies, among other changes (many of which are endogenous to peace),
which make it difficult to test the theory directly.  
This provides motivation for developing a model such as ours that can help sort out the various factors.

Thus, let us now present the model and analyze its implications.  We then conclude with some discussion of other issues such as
the `Democratic Peace' or collective international organizations.

\section{The Model}
\label{sec:model}

\subsection{Countries and Networks}
\label{sub:notation}

There is a set $N=\{1,\ldots, n\}$ of countries.

Countries are linked through {\sl alliances}, represented by a network of alliances $g\subset N^2$ with the interpretation that if $ij\in g$ countries $i$ and $j$ are allies.\footnote{Here we represent a network by the list of unordered pairs $ij$ that it comprises.  So, for instance, the network $g=\{12,23,45\}$ on a set of countries $N=\{1,2,3,4,5\}$ represents situations where country 2 is allied to both 1 and 3, and 4 is allied with 5, and where no other alliances are present.  }
Alliances represent channels through which countries {\sl can} coordinate military actions, either offensively or defensively.
The presence of alliances does not require countries to come to each other's aid, as that will have to be incentive compatible, as embodied in
our definitions below.  The operative part of the assumption is that countries either need to have an alliance or add one in order to coordinate their military activities.  (For more discussion on alliances, see Section \ref{allies}.)

$N_i(g)=\{j: ij\in g\}$ are the allies of $i$.

For a given alliance $ij\notin g$, let $g+ij$ denote the network obtained by adding the alliance $ij$ to $g$.
Similarly, given an alliance $ij\in g$, let $g-ij$ denote the network obtained by deleting the alliance $ij$ from $g$. In a slight abuse of notation, let $g-i$ denote the network obtained by deleting all alliances of the form $ik$, $k\in N$, from $g$; that is, removing $i$ from the network.

Let
$$\mathcal{C} (g) = \{ C\subset N \ | \ ij\in g {\rm \  for \  all \ } i, j \in C\}$$
denote the set of all cliques in a network $g$: that is the set of all groups of countries such that every pair of countries in the group are allied.

\subsection{Military Strengths and Wars}

Each group of countries $C\subset N$ has a collective
military strength $M(C)$.

A prominent example is in which $M(C)=\sum_{i\in C} M_i$ where we write $M_i$ to denote $M(\{i\})$.\footnote{Although it would be interesting to endogenize the strengths, for our purposes in this paper we take these as given. For bilateral models of endogenous military strengths see \citet*{baliga04,jackson09}.}

If there is a war between sets of countries $C$ and $C'$, with $C$ being the aggressor,
 then $C$ ``wins'' if
 $$M(C) > \gamma(C,C') M(C').$$
The parameter $\gamma(C,C')>0$ is the defensive (if $\gamma(C,C')>1$) or offensive (if $\gamma(C,C')<1$) advantage in the war.

The dependence of the parameter $\gamma(C,C')$ on the groups of countries in question allows the model to incorporate various geographic and technological considerations (e.g., land and sea accessibility between countries, nuclear versus conventional capabilities, troop locations, etc.).\footnote{The specification is somewhat redundant at this point since one can incorporate everything into the $\gamma$ function, but this representation will be useful when we specialize it below.}

We maintain very weak monotonicity conditions on the functions:
\begin{itemize}
\item $M(C")\geq M(C)$ whenever $C\subset C"$: bigger groups of countries in terms of set inclusion are at least as strong as smaller groups.
\item $\gamma(C'',C''')\leq \gamma(C,C')$ whenever $C\subset C''$ and $C'''\subset C'$: adding countries to the attacking group and/or subtracting them from the defending group does not increase the defensive advantage.
\end{itemize}

This modeling of a war outcome based on relative strengths is reminiscent of the approach of \citet*{niou1986,niou1991}.  One could instead work with contest success functions (e.g., as in \citet*{skaperdas94,jackson09}), which would provide for random chances of winning.  In our model it would not add anything since we are not focused on arming, and so all that would matter is whether the expected benefits computed via a probability of winning exceed a threshold of potential costs/losses, and so the decisions would still be either to attack or not based on relative strengths and costs and benefits, exactly as already in our model, simply with a different functional form.

\subsection{Vulnerable Countries and Networks}

We say that a country $i$ is {\sl vulnerable} at a network $g$ if there exists $j$ and
$C\subset  N_j(g)\cup \{j\}$ such that $j\in C$, $i\notin C$ and
$$M(C) > \gamma(C, C') M(C'),$$
where $C'=i\cup (N_i(g) \cap C^c)$ and $C^c$ is the complement of $C$.
In this case, we say that country $j$ is {\sl a potential aggressor} at a network $g$.\footnote{A country can be both vulnerable and a potential aggressor at some networks.}

Thus, no country is vulnerable at a network $g$ if
for any coalition $C$ of a potential aggressor $j$ and some its allies, and any target country $i\notin C$,
the aggressors cannot successfully attack the country:
$M(C) \leq \gamma(C,C') M(C')$ where $C' = i\cup (N_i(g) \cap C^c)$.

The incentives of countries to attack or defend are embodied in the definitions below.  The above definitions just define the technology of war.

If some country is vulnerable, then a group that can defeat the country and its remaining allies has an incentive to attack and defeat the country.  This presumes that the benefits from defeating a country outweigh costs of war.  If a country that is not vulnerable were to be attacked then it and its allies would
 be successful in holding off the attackers.  Implicit in the definition is that if the country and its allies could be successful in fending off an attack, then
they would do so.  For now, we simply assume that winning a war (even successfully aiding an ally in defense) is desired and losing a war is not.  When we explicitly model trade and economics below, we will be more explicit about gains and losses.

\subsection{Alternative Definitions of Vulnerability}\label{CC}

In the above definition, in order for a group of countries to attack they must coordinate
via some country $j$, and then the target country $i$ is defended by its neighbors.

We refer to this as `NN-vulnerability'  since the attacking and defending countries can each receive aid from their allies or `neighbors' (hence the `N'), without any additional restrictions.
Of course, we can also consider other definitions.
For example, if we think of alliances as a channel of communication between countries, then
it could be in some circumstances that greater coordination is needed.  For instance, to initiate
an attack, it might be that all countries need to be in communication with all of the others in
the coalition:  i.e., they must form a clique.
Moreover, there could be some asymmetries in military operations.
For example, it might be in some circumstances that attacking coalitions need to be cliques, while a country can
be defended by all neighbors without requiring that the defending coalition be a clique.  This would capture the fact that more coordination is needed when attacking, while defense might only require each neighbor to lend aid to the attacked country. We refer to this case as `CN-vulnerability' (attacking coalitions as `C'liques, defending coalitions as `N'eighbors).

\begin{figure}
	\centering
	\includegraphics[height=2.7in]{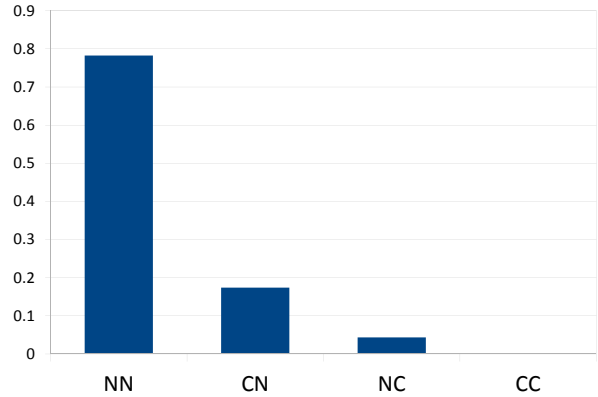}
	\caption{Categorization of wars by type (COW data set from 1823 to 2003 such that at least one side had at least three countries, so some
possibility of a clique).  C indicates that the coalition was a clique (three or more countries with all possible alliances present), while N indicates that the coalition was missing some alliances or involved fewer than three countries.  CN indicates that the ``offense'' (side initiating the conflict according to the COW data classification) was a clique and the ``defense'' (other side) was not a clique.  NN indicates that neither side was a clique, and so forth. }
	\label{figure:NN-CN}
\end{figure}

This is mostly an empirical question, and so let us examine the data on this issue.
Indeed, the fraction of links that are present is higher among attacking coalitions than defending ones.
When considering wars between 1823 and 2003 (from the Franco-Prussian War through the Invasion of Iraq, based on what is available from the Correlates of War Inter-State War database [``COW''] intersected with the ATOP alliance data), 61 percent of the links in attacking coalitions were present while 33 percent of the links in defending coalitions were present (out of 95 wars).
In terms of actual clique counts,
Figure \ref{figure:NN-CN} shows the fraction of wars between 1823 and 2003 that fall into various
categories in terms of whether the attacking/defending coalition was a `C' (clique) or `N' (non-clique - generally a country and some of its neighbors, not forming a full clique, or having fewer than three members).

Figure \ref{figure:NN-CN} shows that the wars predominately involve NN, so fall under our current definition, while a few fall in the categories CN (offense=Clique, defense=Non-clique) and NC, but none in the category CC.
Thus, we focus on the NN-vulnerability definition, but we also provide results for the CN and NC cases, and simply comment on the CC case. In general, we will take the `NN' prefix as understood and simply refer to `vulnerability'.

As we mentioned in the introduction, this diagram also shows that multilateral wars exhibit `network' patterns and do not  fall along the lines of (even subsets) of some partition of countries into coalitions.  This reiterates the motivation for a network-based approach, that becomes even more paramount when we come to the connections with international trade, which exhibits rich network patterns.

\subsection{Illustrations of Vulnerability}

Before moving on to the main definitions and analysis, we present a
simple observation and some illustrations of networks and vulnerability.

\begin{figure}[!h]
\centering
\subfloat[A Network of Alliances]{
	\includegraphics[scale = 0.40]{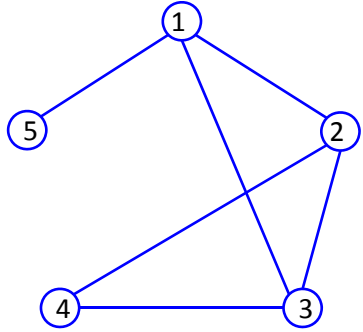}
}\ \ \ \ \
\subfloat[2 and its allies 3 and 4 attack 1 who is defended by 5]{
	\includegraphics[scale = 0.40]{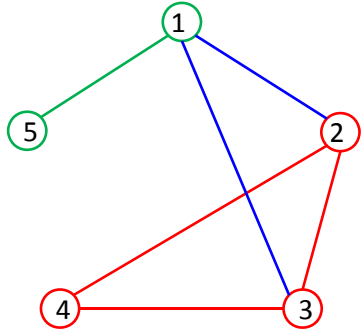}
}
	\caption{1 is vulnerable if $M(\{2,3,4\})> \gamma (\{2,3,4\},\{1,5\})  M(\{1,5\})$.}
\label{figure:example1}
\end{figure}

For an illustration of our definition of vulnerability, consider Figure \ref{figure:example1}.

In this network, country 1 is vulnerable if $M(\{2,3,4\})> \gamma (\{2,3,4\},\{1,5\})  M(\{1,5\})$.  Countries 2,3, and 4 form a clique and hence can attack country 1
under either requirement of C or N for attacking coalitions, and country 1 has country 5 to defend it
under either definition C or N.   Country 5 cannot join countries 2, 3, and 4 in attacking country 1 since it is not allied with any of them.


\begin{observation}
\label{intermediate}
If $M(N\setminus \{i\})>\gamma (N\setminus \{i\}, \{i\}) {M}_i$ for any $i$,
then some country is vulnerable in the complete network.\footnote{I.e.,
the network containing all possible alliances.}
If $M_j> \gamma(\{j\},\{i\}) M_i$ for some $i$ and $j$, then any network which has no vulnerable countries is nonempty and incomplete.
\end{observation}

This simple observation points out that in most settings of interest (i.e., such that a world with no alliances would have some country vulnerable),  requiring that no country be vulnerable implies that networks must be ``intermediate.''

\subsection{Incentives and War-Stable Networks}\label{sub:is}

We now introduce the concept of {\sl war-stability} that accounts for countries' incentives to conquer other countries and to add
or delete alliances.  At this point we do not include trade,
focusing on the world in which alliances serve only military purposes,
and the motivations
for going to war are to gain land or resources from other countries.

The motivation for attacking another country comes from the economic spoils - which historically have included land, natural resources, slaves, and access to markets and other economic resources.
Netted from this are expected damages and other costs of war.
The expected net gain from winning a war is then represented as $E_{ik}(g,C)$, which are the total economic gains that accrue to country $k$ if country $i$ is conquered by
a coalition $C$ with $k\in C$ when the network is $g$ and $i$ is
defended by the coalition $C'=\{i\}\cup N_i(g)\cap C^c$).\footnote{We
allow for the dependence upon the network $g$, since once we allow
for trade, the economic spoils available will be a function of the network.}  For example these include natural resources or other potential spoils of war.\footnote{For important discussions of the spoils of inter-state wars, see \citet*{caselli12,garfinkelss2014}.}


Finally, there are costs to maintaining alliances.  The cost of country $i$ having an alliance with country $j$ is some $c_{ij}(g)>0$.  This could include costs of
opening diplomatic, military and communication channels, coordinating military operations or intelligence, or other related costs.
We generally take costs of alliances to be small relative to the potential spoils of winning a war, as otherwise the analysis is degenerate.
\footnote{In particular,
costs are small enough so that if there
is some $g$ and $jk\notin g$ such that $j$ is a potential aggressor at $g+jk$, but not
at $g$, with $i$ being vulnerable to being conquered by $j$,
then
$c_{jk}(g+jk) + \sum_{s\in N_g(j)}[c_{js}(g+jk)-c_{js}(g)] \leq   E_{ij}(g+jk,C)$.  Thus, $j$ is always willing to add an
alliance to some $k$ that would be pivotal in winning a war.
} The costs are also sufficiently small that any country $i$ is
willing to pay $c_{ij}(g+ij)$ to add an alliance with $j$,
provided that the addition of the alliance would change $i$ from being vulnerable to not.

Define a network $g$ to be \textit{war-stable} if three conditions are met:
\begin{enumerate}
\item[{\bf S1}] no country is vulnerable at $g$;
\item[{\bf S2}] no two countries both benefit by adding an alliance to $g$; and
\item[{\bf S3}] no country has an incentive to delete any of its alliances.
\end{enumerate}

Given that no country is vulnerable at $g$, the only way in which two countries $j$ and $k$ could have an incentive to add a link to $g$ would be that
some other country $i$ must become vulnerable at $g+jk$, and both $j$ and $k$ would have to be part of the winning coalition.  The change in payoffs to $j$ (with an analogous expression for $k$) would be at least $E_{ij}(g+jk,C)- c_{jk}(g+jk) - \sum_{s\in N_g(j)}[c_{js}(g+jk)-c_{js}(g)]$, with $j$ being part of a coalition $C$ that includes $k$ and conquers some $i$.
By assumption, this is positive.  Thus, [{\bf S2}] is equivalent to saying that no country is vulnerable at $g+jk$, $\forall jk\notin g$.
Similarly, given that links are costly, a country not having an incentive to delete any of its alliances implies that
it must be that by deleting any alliance a country becomes vulnerable at the new network.  Therefore [{\bf S3}] is equivalent to saying that both $j$ and $k$
are vulnerable at $g-jk$, $\forall jk\in g$

So, $g$ is \textit{war-stable} if three conditions are met:
\begin{enumerate}
\item[{\bf S1}] no country is vulnerable at $g$;
\item[{\bf S2}] $\forall jk\notin g$, no country is vulnerable at $g+jk$,
\item[{\bf S3}] $\forall jk\in g$,  both $j$ and $k$ are vulnerable at $g-jk$.
\end{enumerate}
That is, $g$ is war-stable if no pair of countries can destabilize the network by adding an alliance and making some other country vulnerable and there are no superfluous alliances.

This definition is similar to that of pairwise stability of \citet*{jacksonw1996} in that we consider changes in the network one alliance at a time, and both additions or deletions - requiring two countries to benefit to form an alliance, but only one country to benefit to break an alliance.  One can enrich the definition in various directions, by allowing groups of countries to add alliances, countries to delete multiple alliances, payments for forming links, forward-forward looking countries, and so forth.  Given that there is a
already a large literature on possible variations on definitions of network formation, we focus on this base definition here.\footnote{See \citet*{jackson2008,blochj2006} for overviews of alternative network formation definitions.}

As a reminder, note that for the definition of `war stability', we use `NN-vulnerability'. When another notion of vulnerability is used then we explicitly note this in the name (e.g., `CN-war-stability' refers to `CN-vulnerability').

\subsection{War and Trade Stable Networks}

We now generalize the model to include payoffs that accrue to countries as a function of the network as a function of trade.

A country $i$ gets a payoff or utility from the network $g$ given by $u_i(g)$.
This represents the economic benefits from
the trade that occurs in the network $g$.

\subsubsection{Alliances}\label{allies}

A link now represents a potential trading relationship and potential to coordinate military activities.
The important assumption is that if two countries trade (significantly) with each other, then they can come
to each other's aid in the event of a military conflict.

The two assumptions that we are using are thus:  (i) having an ``alliance'' involves some costs, however tiny, which must be offset by some benefits either via trade or war, and (ii) without having any relationship, countries are not able to coordinate either in attacking or in defending.

Clearly, (ii) is a simple assumption that `alliances' have some meaning, otherwise there is really nothing to model.  We do allow the formation of new alliances in cases in which that would benefit the countries, as part of the definition.
Alliances can be fairly inexpensive, but still serve a purpose of making
clear who could defend whom in various situations.
One could imagine a more complicated model in which there is some
incomplete information, and in which making public a non-binding alliance
is useful.\footnote{Such a rationale for formal alliances
can be found in, e.g., \citet*{morrow2000}.}

Let us also comment on our presumption that substantial trade
allows for potential military coordination {\sl regardless of whether
there is then an explicit military alliance or not}.  This captures the idea that both the interests and channels of communication are then generally present.
For example, this was exactly what happened in the U.S.
aid to Kuwait in the Persian Gulf War.  Moreover, even though China and the U.S. do not have explicit trading relationships, it would be difficult to imagine the U.S. not reacting if there was an unprovoked attack by some other country on China which severely disrupted trade with the U.S.

This also fits with the empirical evidence.  \citet*{mansfield97} examine correlations between alliances, trade, and participation in preferential trading agreements over the period of 1960 to 1990. They find that alliances (and participation in a preferential trading agreement) lead to increased bilateral trade, with the effect being considerably larger when the pair of countries have both an alliance and mutual participation in a preferential trading agreement.  Interestingly, this relationship differs in the recent 1960-1990 period
compared to pre-World War II.   \citet*{long06}, looking at pre-World War II Europe, finds that trade between allies is only statistically larger than trade between non-allied countries when economic provisions are explicitly mentioned in the alliance.  This fact is consistent with our analysis in that opportunities for trade were substantially limited in pre-World War II Europe, and so the economic trade incentives emerge to a much greater extent in the 1950s and thereafter when costs of trade begin to plummet and incomes increase and trade grows significantly.   Regardless of the relationships between {\sl explicit} alliances and trade, the open lines of communication are what is essential for our theory.

\subsubsection{Vulnerability and Stability with Trade}

We now introduce a concept of vulnerability based on the incentives of countries to attack others when explicitly accounting for the benefits and costs associated with conquering a country.


We say that a country $i$ is {\sl vulnerable despite trade} in a network $g$
to a country $j$ and coalition $C\subset N_j(g)\cup\{j\}$ if $j\in C$, $i\notin C$ and
\begin{itemize}
\item $M(C)>\gamma(C,C') M(C')$ where $C'=i\cup(N_i(g)\cap C^c))$  (i.e., $C$ could conquer $i$), and

\item $u_k(g-i)+E_{ik}(g,C) \geq u_k(g)\;\forall k\in C$ with some strict inequality: every $k\in C$ would benefit from conquering $i$, factoring in economic gains of conquering and gains or losses in subsequent payoffs from the network.\footnote{It is not essential whether the strict inequality is required for all countries or just some, or must include $j$, as generically in the $E$ function ensures there will not be equality for any countries.}
\end{itemize}

The second item is new to this definition of vulnerability and incorporates two aspects of economic incentives of countries to attack each other:

Recall that the $E_{ik}(g,C)$ represents the potential net benefits that $k$ enjoys from conquering $i$ as part of the coalition $C$ in a network $g$.
If a country is poor in natural resources, and much of its economy is built upon nontransferable or difficult to extract human capital, it would tend to have a lower $E_{ik}$ and would be less attractive.

The $u_k(g-i)$ accounts for the fact that as $i$ is conquered then the network of trade will adjust.  If $k$ is a trading partner of $i$, then $k$ could lose via the elimination of $i$, with $u_k(g-i)< u_k(g)$.\footnote{As \citet*{glick10} documents, the economic loss resulting from trade disruption during wars can be of the same order as more traditional estimates of losses resulting from interstate conflict.  This does not even account for the potential loss of trade if a partner is lost altogether.}  Note that this effect works both ways: it might also be that a country $k$ benefits from the elimination of some country $i$, for instance if it improves $k$'s position in the resulting trade network.

With this framework, we now define a stability notion corresponding to war stability but adding the economic considerations.

Our definition of war and trade stability now incorporates two incentives for adding or deleting alliances.  First, countries might add or maintain an alliance because of its military value in either preventing a war or in instigating one, just as with war stability.  This is similar to what we considered before, except that countries now consider the economic spoils and consequences of war
in deciding whether to take part in an attack.
Second, countries add or maintain alliances for the economic benefits in terms of trade.

Let us now consider the incentives for countries to add an alliance and attack another country.

Starting from a network $g$, some alliance $jk\notin g$ is {\sl war-beneficial} if
there exists some $C\in N_j(g+jk)\cup\{j\}$ with $j\in C$, $k\in C$ and $i\notin C$ such that $i$ is vulnerable despite trade to $C$ at $g+jk$ and
\begin{itemize}
\item $u_j(g+jk-i)+E_{ij}(g,C)\geq u_j(g)$, so, $j$ would benefit from forming the link and attacking, and
\item $u_k(g+jk-i)+E_{ik}(g,C)\geq u_k(g)$, similarly for $k$, with one of these inequalities holding strictly.
\end{itemize}

We say that a network $g$ is {\sl war and trade stable}  if three conditions are met:
\begin{itemize}
\item[{\bf ES1}] no country is vulnerable despite trade at $g$;
\item[{\bf ES2}] $\forall jk\notin g$: if $u_j(g+jk)> u_i(g)$ then $u_k(g+jk)< u_k(g)$, and also $jk$ is not war-beneficial
\item[{\bf ES3}] $\forall jk\in g$ either $u_j(g-jk)\leq u_j(g)$ or $ j$ is vulnerable despite trade at $g-jk$, and similarly for $k$.
\end{itemize}

So, a network of alliances is war and trade stable if no country is vulnerable despite trade, if no two countries can add an alliance and both profit either through economic or war means, and either economic or war considerations prevent any country from severing any of its alliances.

\begin{remark} If $u_i(\cdot)$ is constant for all $i$, then  war  and trade stability
reduces to war stability.
\end{remark}

We say that a network $g$ is {\sl strongly war and trade stable} if it is {\sl war and trade stable} for any (nonnegative) specification of the $E_{ij}$'s.

Again, the default definition will be relative to a country and its neighbors attacking or defending, but the same extensions to cliques hold as in the earlier sections (so, there are CN, NC, and CC variations on the definitions).  The default definition refers to the NN case.

We remark that we have not explicitly mentioned the incentives of a country that loses a war.  Clearly, there are costs to losing, and what they are exactly does not matter in the model.   Being explicit about the losing countries' payoffs is not necessary, since our stability notions only need to check whether countries would benefit by attacking, benefit by adding an alliance and then attacking, or could safely remove an alliance; and the payoff to the losers is irrelevant in these calculations.

\section{Nonexistence of War-Stable Networks}
\label{subsub:isresults}

For the case of 2 countries, it is direct to check that the only possible stable network is the empty network and it is war-stable only if each country has a sufficient defensive advantage.
Thus, we consider the more interesting case with $n\geq 3$.

Before presenting the results on lack of existence of war-stable networks, let us illustrate some of the main insights.


We start with a very simple network that is never war-stable: the complete network. 
First, in order for no country to be vulnerable it is clear from Observation \ref{intermediate} that it would have to be that $M(N\setminus \{i\})\leq \gamma(N\setminus \{i\} , \{i\})  M_i$ for all $i$.
However, in that case any country $i$ can delete any of its alliances and still not be vulnerable, violating the condition for war stability.

\begin{figure}
	\centering
	\includegraphics[height=3.0in]{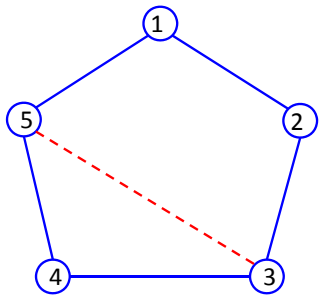}
	\caption{A network that is \emph{not} war-stable for
any parameter values. }
	\label{figure:RingUnstable}
\end{figure}

The argument is slightly different, but the same conclusion applies to less connected networks, as in the network pictured in Figure \ref{figure:RingUnstable}.  Begin with the ring network in which each country has two links.
Let us examine the NN case (variations on the argument apply in other cases).\footnote{For the cases of CN and CC, the argument is more involved (see Claims \ref{c00} and \ref{c0} for details). }  In order for 1 not to be vulnerable under the addition of the link 53, it must be that $\gamma( \{2,3,4,5\},\{1\}) M_1\geq M(\{2,3,4,5\})$ (as it must not be vulnerable to 3 and its allies 2,4,5).  However, this implies that 1 is not vulnerable in the original network if it deletes an alliance regardless of the attacking coalition, and so this contradicts war-stability.

The following theorem shows that there are no war-stable networks except empty networks in extreme cases, regardless of country's strengths.
This particular theorem applies under NN-vulnerability.


\begin{theorem}
\label{IS-NN}
Let $n\geq 3$. There are no nonempty war-stable networks.
The empty network is war-stable if and only if  $M(\{j,k\})\leq \gamma(\{j,k\},\{i\}) M_i$ for all distinct $i,j,k$.
\end{theorem}

Theorem \ref{IS-NN} shows that war-stable networks only exist in the extreme case in which the defensive parameter is so high
that the weakest country can withstand an attack by the two strongest countries in the world, in which case the empty network is stable.  Outside of that case, there are no war-stable networks.
The intuition behind the proof of Theorem \ref{IS-NN} is similar to that of the examples: outside of the extreme case, requiring that
a country not be vulnerable, nor vulnerable to the addition of any alliances, implies that a country has extraneous alliances.

The nonexistence of war-stable networks extends to other definitions of vulnerability as we now verify.

\begin{theorem}
\label{IS-Robust}
Let $n\geq 3$.
\begin{itemize}
\item {\bf NC-vulnerability:} There are no nonempty NC-war-stable networks.
The empty network is NC-war-stable if and only if $M(\{j,k\})\leq \gamma(\{j,k\},\{i\}) M_i$ for all distinct $i,j,k$.
\item {\bf CN-vulnerability:} Under the uniform strength case of $M_i=\bar{M}\;\forall i$ and $\gamma(C,C')=\gamma$ for all $C,C'$, if $1\leq \gamma <2$ then there are no CN-war-stable networks.
If $\gamma\geq 2$, then the unique CN-war-stable network is the empty network.
If $\gamma<1$, then for large enough $n$, there exist nonempty CN-war-stable networks.
\end{itemize}
\end{theorem}

Even though the arguments for any particular network's instability are straightforward, showing that there do not exist any
nonempty war-stable networks under these variety of definitions requires covering all possible configurations, and so is quite involved.
Thus, the full proof of the theorems, including the case of CN-vulnerability, uses a combinatorial pigeonhole argument, showing that certain sorts of contradictions arise in all nonempty graphs.
The case of CN-vulnerability is quite intricate, and so our proof is limited to the case of equal strengths, and we are unsure of
the full characterization for the asymmetric case.\footnote{\label{footnote}
For CN-war-stability, the restriction to $\gamma\geq 1$ is important.  If the offense has a substantial advantage and $\gamma<1$, for the case
 of CN-vulnerability there exist war-stable networks.  At first blush it might be surprising
 that a world where attackers have an advantage over defenders leads to more stability, but it can be understood as follows.  An offensive advantage provides incentives for
 countries to maintain alliances, as without alliances countries easily become vulnerable. This allows one to build up networks of alliances that are denser.  The key to then getting CN-war-stability
 is to have each country be involved in several separate cliques, so that no attacking clique is large enough to overcome the country and its other allies (see Section \ref{CN-vuln}).
 The case of CC-war stability is particularly challenging.  We can show nonexistence when $\gamma<1$, and conjecture that it also holds for $\gamma>4/3$, but can find some CC-war-stable networks for $1\leq \gamma\leq 4/3$.  Given that it is not a case of empirical interest, we leave it aside.}

\subsection{Nuclear Weapons}

An obvious trend that occurs post WWII is that nuclear weapons were invented during the war and greatly enhanced
in both power and delivery methods through the following decades, leading to dramatic changes in the
technology of war.  Although rarely used, their existence changes the technology and potentially the opportunities for stability.\footnote{There
is a large literature on the cold war and a contentious debate on the potential stabilizing or destabilizing impact of nuclear technology (e.g., see \citet{schelling1966,mueller1988,geller1990}.}

We emphasize that their existence alone does not lead to stability: our model (for the most relevant NN case, as well as the
NC case) allows for completely arbitrary
asymmetries in military strengths and in offensive/defensive advantages.  There is no way for countries to ally themselves, as a function of their strengths and nuclear capabilities, to produce a stable and nonempty network.
The only way in which nuclear weapons could stabilize things would be for all countries to have them and for the empty network to ensue.
This is clearly not the case.

Therefore, the model makes clear that one needs trade, or some other consideration, to explain why we see denser networks that are stable and why many non-nuclear countries also live in relative peace.   Thus, we now turn to the analysis of the stability when there are substantial trade considerations.

\section{Existence of War and Trade Stable Networks}
\label{sub:ies}

As we have seen, pure military considerations do not lead to stable networks.
As we show next, accounting for economic incentives associated with gains from trade can restore stability.

\subsection{Results on, and Examples of, War and Trade Stable Networks}
\label{sub:iesresults}

Let us examine the set of war and trade stable networks.
We begin by identifying a condition that is sufficient for war and trade stability.

\begin{proposition}
\label{pairwisesuff}
Suppose that $g$ is pairwise stable with respect to $u$.  If no country is vulnerable despite trade at $g$ or $g+jk$ for any $jk\notin g$, then $g$ is war and trade stable. Moreover, if no country is vulnerable at $g$ or $g+jk$ for any $jk\notin g$,
then $g$ is  strongly war and trade stable.
\end{proposition}

The proof of the proposition is straightforward and thus omitted (and also extends to the CN, NC, CC definitions).

There are many examples of networks that are war and trade stable but not war stable.  The following theorem outlines a whole class of war and trade stable networks, showing that economic considerations restore general existence results.

For what remains, which are constructive results, we specialize to the case of symmetric countries (so the $u_i(\cdot)$, $E_{ij}(\cdot)$, and $M_i$'s are independent of $i$ and $j$), but it will be clear that similar results extend to the asymmetric case with messier statements of conditions.

We also consider a canonical case in which
$$u_i(g)=f(d_i(g))-c\cdot d_i(g),$$
where $d_i(g)$ is the degree of $i$ and $f$ is concave, nondecreasing, and such that there exists some $d\leq n-1$ such that $f(d)<c\cdot d$.  This is a simple model of gains from trade and costs of having trading relationships, abstracting from heterogeneity in goods and trading partners and inter-dependencies in trading relationships beyond diminishing returns - but illustrates our main point and it should be clear that similar results hold for richer models.  Let $d^*$ maximize $f(d)-c\cdot d$ among nonnegative integers.

In addition, in this model and given the symmetry, let $E_{ij}(g,C)= \frac{E(d_i(g))}{|C|}$,  so that each country's economic spoils from a war depend only on that country's degree, and then are divided equally among the attacking countries.

\begin{theorem}
\label{econstable}
Consider the symmetric model with $d^*\geq 2$.
\begin{itemize}
\item Any $d^*$-regular network (i.e., such that each country has $d^*$ alliances) for which no two countries have more than $k<d^*-1$ allies in common is strongly war and trade stable network if $\gamma \geq \frac{d^*+1}{d^*-k-1}$.
\item If   $E (d^*)  \leq  2[f(d^*) - f(d^*-1) - c]$,
then any $d^*$-regular network (in any configuration, including combinations of cliques) is war and trade stable network if $\gamma \geq \frac{d^*+1}{d^*-1}$.
\end{itemize}
\end{theorem}

Theorem \ref{econstable} provides two existence results that each work from a different idea.
The first part is based on the motivation that trade provides for countries to maintain relationships for trade purposes, and the fact
that this results in networks in which no country is vulnerable, and no country would be vulnerable even with the addition of new alliances.  This first result is independent of the $E_{ij}$'s and the relative costs of war, but does require some specific structures (for example, simply forming cliques where each country has $d^*$ allies will not work, as then all of a country's partners can attack the
country and win).
The second result applies for more specific gains from war ($E_{ij}$s), but for a wider set of networks.  It works off of the fact that with sufficient gains from trade, the potential spoils of a war are outweighed by the lost trade value - and so countries are never attacked by one of their own trading partners.  In that case, each country then has enough alliances to protect itself against attacks from outside, and then a wide range of networks becomes stable.   This allows for more cliquish structures to be stable, which are more consistent with the emerging networks that we observe in the world today.
Thus, we see two different ways in which trade stabilizes the world.

\begin{figure}[!h]
	\centering
	\includegraphics[angle=270, scale=0.35]{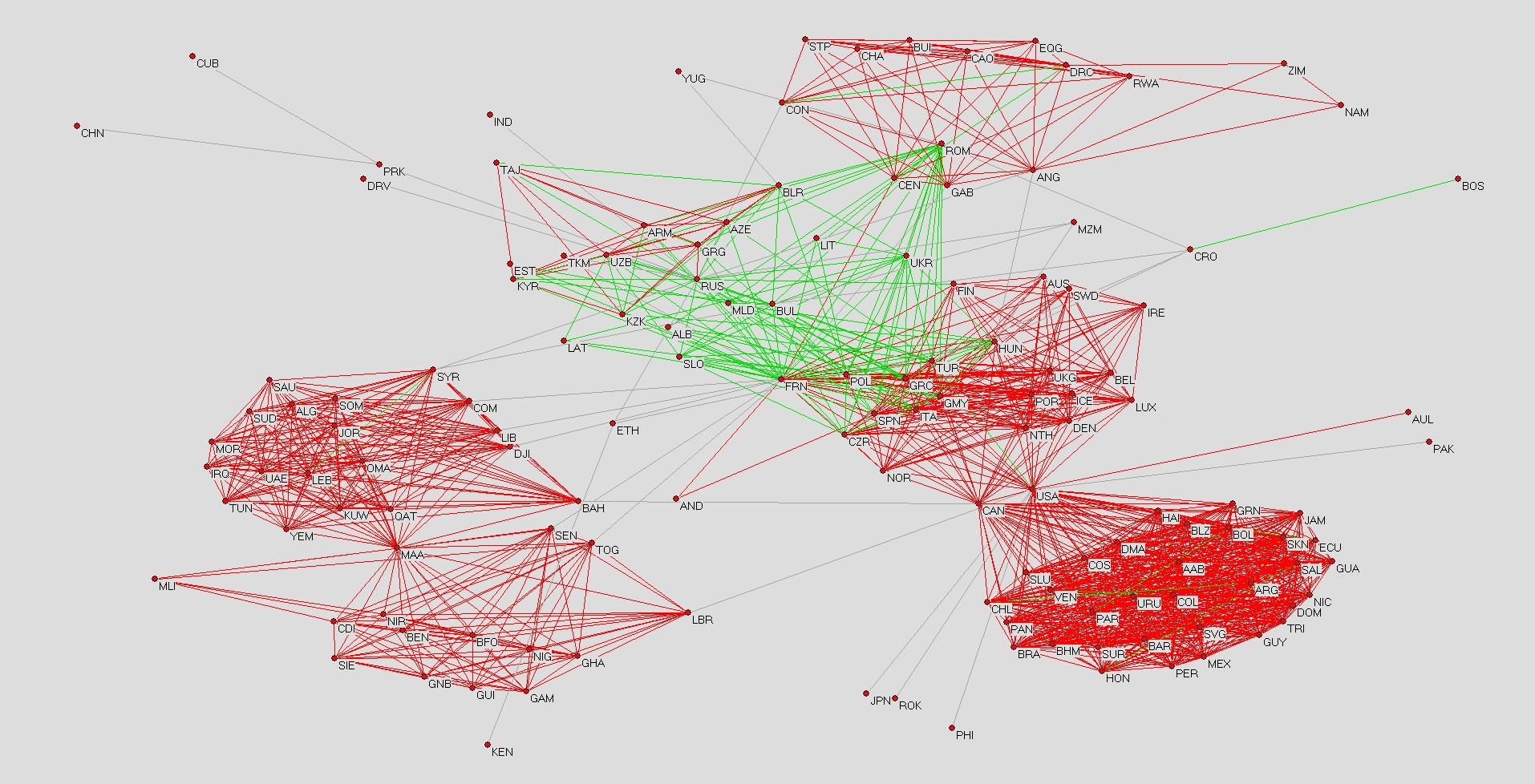}
	\caption{Network of Alliances, 2000, red for multilateral alliance, grey for bilateral alliance, green for both}
	\label{figure:2000}
\end{figure}

Variations on the result hold for other definitions.
For the case of CN-war-and-trade stability, if $d^*\geq 4$ then any network in which all countries have $d^*$ alliances, the largest clique is
of size at most $\lceil \frac{d^*}{2}\rceil$, and any two cliques intersect in at most one country is CN-strongly war and trade stable for any $\gamma\geq 1$.\footnote{\label{econstable1}
A particularly interesting class of CN-strongly war and trade stable networks is one that is built up from a set of cliques, called quilts.
A network is said to be a {\sl quilt} if all nodes have at least two links and the network can be written as a union of cliques, each of size at least 3, and such that any two cliques share at most one node in common.  (This definition of quilts differs slightly from that of social quilts introduced by \citet*{jackson12} in that larger cycles are permitted here.)
In particular, if $d^*\geq 4$ then any quilt in which all countries have $d^*$ alliances and the largest clique is
of size at most $\lceil \frac{d^*}{2}\rceil$ is strongly war and trade stable for any $\gamma\geq 1$.
Quilts are of interest as we see their underpinnings in, for example, the network in Figure \ref{figure:2000}, consists of a number of cliques that have some small overlap.  With $d^*=3$, the same proposition holds with $\gamma\geq 3/2$ and with $d^*=2$ it then moves to $\gamma\geq 2$. }

\section{A Second Look at Data}

The model suggests some dimensions that are important in determining peace that were not investigated in previous studies.  For instance, as just discussed, a country having more allies (who are trading partners) and having more trade with them would lead the country to be less prone to attack.  Also, the country should be less prone to being attacked by countries that trade with it substantially.
Before closing, we take a brief look at these effects in the data, and we concentrate on the period of  1950-2000 for which we have detailed trade and GDP data  (from Gelditsch (2002)).

Table \ref{dyad_trade_conflict} reports a logistic regression of the probability that two countries are at war with each other in a given year as a function of the
level of trade between the two counties, where the level of trade between the two countries is a measure of the total exchanged (imports plus exports) divided by the maximum of the GDPs of the two
countries as a normalizer.  We limit attention to countries within 1000 km of each other as most other dyads are much less likely to be at war or trade with each other.
We also consider war to be a MID 5 according to the COW data set.

\begin{table}[htbp]
\def\sym#1{\ifmmode^{#1}\else\(^{#1}\)\fi}
\begin{center}
\caption{Logistic Regressions of Dyadic War on Dyadic Trade\label{dyad_trade_conflict}}
\begin{tabular}{lcccc}
\hline\hline
                    &(1)	&(2)		&(3)  &(4)	\\
\hline
Dyad Trade   &      -1974.37\sym{***}  &  &   -1974.37\sym{**} &  	 \\
			&        (383.69)   & 	  	      &  (964.74)  &  			\\

Lagged Dyad Trade   &       &  -1150.24\sym{***} &     &  -1150.24  	 \\
												&        	  	  & (248.29)  &     & (758.19)			\\

\hline
Decade Dummies					&				Yes				&				Yes						&			Yes			& Yes   	\\
\hline
Standard Errors Clustered  &   	No & 	No &  Yes & Yes  \\ 
at Dyad Level & & & &  \\						
\hline
Observations        &        36832      & 	 35658 					&        36832       &  35658						\\
\hline\hline
\end{tabular}
\end{center}

{\footnotesize Standard Errors in parentheses. \sym{*} \(p<0.1\), \sym{**} \(p<0.05\), \sym{***} \(p<0.01\)
 Logit regression of dyad in conflict on dyadic trade. Consider dyads within 1000km of each other. Dyad at war if involved on opposite sides of an MID 5. Dyad trade is normalized by the minimum of the two countries' GDPs. Conflict data from COW. Trade and GDP data from Gleditsch (2002). Distance data from Gleditsch and Ward (2001).}
\end{table}

In Table \ref{dyad_trade_conflict} we see a large and significantly negative relationship between the trade between two countries and the chance that they will go to war, and the results are robust to the inclusion of country fixed effects (with clustered standard errors) and controls for the decade.
To get an idea of the magnitude of the effect, a one standard deviation (.0087) increase in the normalized dyadic trade decreases the log odds ratio that two countries are at war with each other by a factor of more than 17 (based on the coefficient from column (1,3)) - basically taking the odds ratio to 0.

We also explore the same relationships, but for the case in which we only look at {\sl new wars} - so the first year that countries are at war.  This controls for the fact that two countries' trade may decrease once they go to war or are at war for a while.  These results are reported in Table \ref{dyad_trade_conflict-bis}.  Again, we see large negative effects.  The significance of the effects again depends on whether standard errors are clustered at the dyad level.

\begin{table}[htbp]
\def\sym#1{\ifmmode^{#1}\else\(^{#1}\)\fi}
\begin{center}
\caption{Logistic Regressions of {\sl New} Dyadic War on Dyadic Trade\label{dyad_trade_conflict-bis}}
\begin{tabular}{lcccc}
\hline\hline
                    &(1)	&(2)		&(3)  &(4)	\\
\hline
Dyad Trade   &      -660.85\sym{***}  &  &   -660.85 &  	 \\
			&        (280.11)   & 	  	  &  (441.48)  &  			\\
Lagged Dyad Trade   &       &  -337.29\sym{**} &     &  -337.29  	 \\
				&        	  	 & (171.34)  &     & (326.12)			\\
\hline
Decade Dummies					&		Yes	&	Yes		&	Yes			& Yes   	\\
\hline
Standard Errors Clustered  &   	No & 	No &  Yes & Yes  \\
at Dyad Level & & & &  \\						
\hline
Observations        &        35565     & 	 35565 					&       35565        &  35565						\\
\hline\hline
\end{tabular}
\end{center}

{\footnotesize Standard Errors in parentheses. \sym{*} \(p<0.1\), \sym{**} \(p<0.05\), \sym{***} \(p<0.01\)
 Logit regression of dyad in conflict on dyadic trade. Consider dyads within 1000km of each other. Dyad at war if involved on opposite sides of an MID 5. Dyad trade is normalized by the minimum of the two countries' GDPs. Conflict data from COW. Trade and GDP data from Gleditsch (2002). Distance data from Gleditsch and Ward (2001).}
\end{table}

Table \ref{country_trade_conflict2} reports a logistic regression of the probability that a country is at war with some other country in a given year as a function of the number of trading partners that the country has (i.e. the number of countries with which the country trades at least .5 percent of its GDP), as well as the total trade that the country has with its partners normalized by its GDP.

\begin{table}[htbp]
\def\sym#1{\ifmmode^{#1}\else\(^{#1}\)\fi}
\begin{center}
\caption{Logistic Regressions of Country being at War based on its Trade with Allies\label{country_trade_conflict2}}
\begin{tabular}{lcccc}
\hline\hline
                    &(1)	&(2)	&(3)	 &(4) \\
\hline
Number Trading Partners &      -0.0797\sym{*}  & -0.0906 &		-&-\\
						&     (0.0476)      & (0.0598) 						&-	&	-\\
[1em]
Normalized Ally Trade   &       -        & 				-			  & 			 -11.299\sym{***}	& -8.936	\\
												&      -          & 	  	-		& (2.795)	& (6.987) \\

\hline
Decade Dummies					&			Yes				&		Yes				&Yes				&Yes			\\												
\hline
Country Fixed Effects   &				No		&				Yes						&   No		& Yes		\\
\hline
Standard Errors Clustered   &				Yes&				Yes			&			Yes			& Yes		\\
at Country level  &   &  &   &    \\
\hline
Observations        &        6760      & 1464									 & 6760			&		1464	\\
\hline\hline
\end{tabular}
\end{center}
{\footnotesize Standard Errors in parentheses. \sym{*} \(p<0.1\), \sym{**} \(p<0.05\), \sym{***} \(p<0.01\). Logit regression of country at war on measures of trade. Country at war if involved in an MID 5 with another country within 1000km. Number of trading partners of country A defined as number of countries with which country A trades at least 0.5\% of its GDP. Country A's trade with allies is normalized by its GDP. Conflict data from COW. Trade and GDP data from Gleditsch (2002). Distance data from Gleditsch and Ward (2001). Ally data from ATOP}
\end{table}

Again, we see significantly negative relationships - although here putting in the large number of fixed effects cuts into sample size and eliminates significance given that we are
clustering standard errors at the country level (although the coefficients are highly significant if we do not cluster the standard errors).
To get an idea of the magnitude of these effect, each added ally decreases the log odds ratio that a country is at war by .08 and so the odds ratio that a country is at war decreases by roughly 8 percent, and adding ten allies (just under the mean) decrease the odds
that a country is at war by over 50 percent.\footnote{Countries are generally at peace, so the odds ratio is close to the probability that a country is at war.}

These relationships are consistent with the model. Moreover, these results suggest that taking a networked view of war, trade, and alliances is a promising avenue for future empirical work.


\section{Concluding Remarks}
\label{sec:concl}

We have provided a first model through which to analyze networks of military alliances and the interactions of
those with international trade.
We have shown that regardless of military technologies and asymmetries among countries, stable networks fail to
exist unless trade considerations are substantial.

Although this points to trade as a {\sl necessary} condition for stability, whether it will be sufficient for stability depends
on how large the benefits are and how large the costs of war are.

In closing, let us comment on several other features of international relations that are part of the larger picture of inter-state war.

A notable change in alliances during the cold-war period was from a `multipolar' to a `bipolar' structure, something which has been extensively discussed in the cold-war literature (e.g., see
\citet*{bloch2012} for references).
Although this lasted for part of the post-war period, and was characterized by a stalemate between the eastern and western blocks, such a system of two competing cliques of alliances is only war
stable if there are sufficient trade benefits between members of a clique, as shown in Theorem \ref{econstable}.
Moreover, it is more of a historical observation than a theory, and it does not account at all for the continued peace that has ensued over the last several decades.  Thus, this fits well within the scope of the model and does not account for the overall trend in peace.

Another institutional observation regarding the post-WWII calm is that institutions have allowed for coordination of countries onto a peaceful ``collective security'' equilibrium where any country disrupting international peace is punished by all other countries, so that war against one is war against all. However, as shown by \citet*{niou1991}, this equilibrium is in some sense ``weak'': it relies heavily upon the assurance that a country tempted to join an attacking coalition will refuse and that all countries will follow through on their punishment commitments, so that far-sighted expectations of off-equilibrium behavior are correct. Given that various small conflicts since WWII did not precipitate a global response, such doubts of some countries' commitment to follow through on punishments seems reasonable.\footnote{One might think of the rise of international institutions as allowing larger groups of countries to simultaneously add alliances, rather than the pairwise addition in our base model below.  However, altering our definition of stability to admit coalitions of countries adding alliances only decreases the set of potentially war-stable networks, once again indicating that trade needs to be incorporated
into a model of alliances in order to account for the dramatic drop in conflict and
simultaneous increase in alliances (strongly correlated with trade).}
Although this then does not seem to explain the lasting peace, it nonetheless does suggest an interesting avenue for extension of our model: taking a repeated games approach to multilateral conflict when including networks of alliances and trade.

One more relevant observation regarding changes in patterns of conflict is the so-called ``Democratic Peace'' : democracies rarely go to war with each other.\footnote{For early background see (\citet*{kant1795,doyle1986,russet1993}) and more recent references can be found in  \citet*{jackson11}.}  This coupled with a large growth of democracies might be thought to explain the increase in peace.
However, once one brings trade back into the picture, it seems that much of the democratic peace may be due to the fact that well-established democracies tend to be better developed and have higher level of trade.
Indeed, studies by \citet*{mousseauho2003} and \citet{mousseau2005} indicate that poor democracies are actually significantly more likely to fight each other than other countries, and that paired democracy is only significantly correlated with peace when the countries involved have high levels of economic development; which is consistent with trade playing the major role rather than the government structure.
Our model abstracts from political considerations, which still could be significant, and so this suggests another avenue for further extension.\footnote{For example, \citet*{concinisz2009} discuss how term limits and electoral accountability affect the incentives to go to war, and \citet*{jacksonm2007} discuss the divergence of the incentives, between politicians and the population that they represent, to go to war.}

We close by noting two other obvious ways in which to enrich our model.
First, one can enrich the modeling of trade.
There are many ways to introduce heterogeneity, for instance along the lines of \citet*{dixit77}; or else, capturing the complexity of trade dynamics as discussed in \citet*{gowa04}, \citet*{long06}, and \citet*{mansfield97}.
Second, and relatedly, is a question of geography.
Both trade and war have strong relationships with geography (see, e.g., \citet*{eaton02}, as well as \citet*{caselli12}, who find that from 1945 to 1987 eighty six percent of significant international wars were between neighboring states).
Geography constrains both the opportunities and benefits from trade and war, and so it has ambiguous effects on stability.
Nonetheless, it plays an important role in explaining realized networks of trade and alliances and deserves further attention.


\nocite{*}

\bibliographystyle{plainnat}
\bibliography{WarNetworks}



\section{Appendix:  Proofs}

\noindent{\bf Proof of Theorems \ref{IS-NN} and \ref{IS-Robust}:}

For any of the definitions of vulnerability (NN, CN, NC, CC), the conditions on stability can be recast as requirements on the $\gamma$ parameter.
Let $\mathcal{C}^*(g)$ denote the feasible attacking coalitions under the corresponding definition - either some $j$ and a subset of its neighbors, or a clique.

Before treating the CN case, we complete the proof for the cases of NN and NC.
Consider a country that has an alliance in a nonempty network, say $i$, which then has alliance to some $k$.
In order for [S3] to be satisfied, it must be that $i$ is vulnerable in $g-ik$.
Thus, there is some $j$ and $C\subset N_{j}(g-ik) \cup {j}$ with
$M(C)> \gamma(C,C') M(C')$ for every feasible $C'$ (depending on the NN or NC case) out of all $C'\subset \{i\}\cup N_i(g-ik)\cap C^c$ that can defend $i$.
Given [S1], it must be that $i$ was not vulnerable at $g$, and so it must be that $k\notin C$ and in particular that $jk\notin g$.
However, if the link $jk$ is added (so that the network $g+jk$ is formed), then $C\cup \{k\}$ can defeat $i$, since $M(C\cup \{k\})\geq M(C)$
and also $\gamma(C\cup \{k\},C') M(C') \geq  \gamma(C,C') M(C')$
for any feasible  $C' \subset \{i\}\cup N_i(g)\cap C^c /setminus \{k\}$ that can defend $i$, and so
$$M(C\cup \{k\})\geq M(C)>\gamma(C,C') M(C')\geq \gamma(C,C') M(C')$$ for any feasible  $C' \subset \{i\}\cup N_i(g)\cap C^c /setminus \{k\}$ that can defend $i$.   This violates [S2] as then $j$ and $k$ benefit from adding the link since $i$ is vulnerable to a coalition containing
both $j$ and $k$, which is a contradiction.
This establishes that any network that is NN or NC-war-stable must be empty.

We now turn to the CN case.
The first condition that no country be vulnerable, [S1], translates as:

\begin{equation}
\label{eqis1}
\gamma\geq\max_{C\in\mathcal{C}^*(g)}(\max_{i\in C^c}\frac{M(C)}{M(i\cup(N_i(g)\cap C^c))}).   \ \ \ {\rm [S1]}
\end{equation}

The second condition that no additional link leads to [S2] translates as:
\begin{equation}
\label{eqis2}
\gamma\geq\max_{jk\notin g}(\max_{C\in\mathcal{C}^*(g+jk)}(\max_{i\in C^c}\frac{M(C)}{M(i\cup(N_i(g+jk)\cap C^c))})).
\end{equation}
Note that given (\ref{eqis1}), we need only check (\ref{eqis2}) with respect to $C$ such that $j\in C$ and $k\in C$.  Thus, we can change the
denominator in (\ref{eqis2}) to be $M(i\cup(N_i(g)\cap C^c))$.
Therefore, stability implies that
\begin{equation}
\label{eqis2b}
\gamma\geq\max_{jk\notin g}(\max_{C\in\mathcal{C}^*(g+jk)}(\max_{i\in C^c}\frac{M(C)}{M(i\cup(N_i(g)\cap C^c))})).  \ \ \ {\rm [S2]}
\end{equation}

The third condition [S3] translates as (providing $g$ is nonempty):
\begin{equation}
\label{eqis3}
\gamma<\min_{ij\in g}(\min\{\max_{C\in\mathcal{C}^*(g-ij)}\frac{M(C)}{M(i\cup(N_i(g-ij)\cap C^c))}, \max_{C\in\mathcal{C}^*(g-ij)}\frac{M(C)}{M(j\cup(N_j(g-ij)\cap C^c))}\}).   \ \ \ {\rm [S3]}
\end{equation}

Label countries in order of decreasing strength, so that $M_i\geq M_{i+1}$.

First, note that the empty network is stable (under any of the vulnerability definitions), if and only if $(M_1+M_2)/M_n \leq \gamma$.
This follows since in that case [S1] is clearly satisfied, and also [S3] is vacuously satisfied since there are no links to delete, and since $(M_1+M_2)/M_n \leq \gamma$ corresponds to the cases where [S2] is satisfied.   Thus, to prove the theorems, it is enough to show that there are no nonempty war-stable networks
for NN or NC; and for CN when $\gamma\geq 1$.

We begin with a claim that applies regardless of the vulnerability definition (NN, CN, NC, CC)

\begin{claim}
\label{c00}
There does not exist a non-empty war stable network with maximum degree less than 2.
\end{claim}

\noindent{ \bf Proof of Claim \ref{c00}:}
Consider a network with a maximum degree of 1.   If $\gamma< 1$, then the network must violate [S1], since a (strongest) country in any linked pair can defeat the other country.
So, consider the case in which $\gamma \geq 1$.
Let $n$ be the weakest country.
Let $i$ either be the ally of $n$, or else some other country if $n$ has no allies.  It follows that $\gamma\geq 2$, as otherwise
$i$, together with some country $k$ different from $i$ and $n$ that is either an existing ally of $i$'s or by forming a new link $ik$, could defeat $n$, which would violate [S1] or [S2] respectively.
However, $\gamma\geq 2$ implies that the network cannot be war stable.
 This is seen as follows.  Consider the strongest country $i$ that has positive degree.
 Either $i$ can sever its link violating [S3], or else (given that $\gamma\geq 2$ and $i$ is the strongest among those having connections and cliques are at most pairs) it must be that there is some country $k$ that has
 no ties that could defeat $i$ if $i$ severed its link to its ally $j$.   However,
 then by adding $ik$ they would defeat $j$ (since $j$ is no stronger than $i$ and $i$ would be defeated
 by $k$ when $k$ is all alone) violating [S3].\eproof

We now specialize to equal strengths for the remainder of the proof which covers the case of CN.

\begin{claim}
\label{c0}
There does not exist a non-empty CN-war stable network with maximum degree less than 3.
\end{claim}

\noindent{ \bf Proof of Claim \ref{c0}:}
Given Claim \ref{c00}, consider a network $g$ with maximum degree two.

First, consider the case in which $\gamma\geq 2$
Given that the biggest clique is of size 3 and $\gamma\geq 2$, then a country $i$ with degree 2 could sever one of its links and not be CN-vulnerable (its remaining ally cannot be part of any clique of size more than 2), and any clique of size 3 could not defeat $i$ and its remaining ally.   Thus there is no country with degree 2 if the maximum degree is 2, which is a contradiction.

So, consider the case in which $\gamma< 2$, and consider a country $i$ and links $ij\in g$ and $ik\in g$.
It cannot be that $jk\in g$ as otherwise $jk$ can defeat $i$, violating [S1].  Similarly, if $jk\notin g$
then by adding that link $jk$ would defeat $i$ violating [S2].  So, again we reach a contradiction.





Thus, it must be that the maximum degree is at least three.\eproof

\begin{claim}
\label{c1}
Consider $i$ of maximum degree and some $ij\in g$.
There exists $C \in \mathcal{C}(g-ij)$ such that
\begin{equation}
\label{eqis3b}
\gamma<\frac{M(C)}{M(i\cup(N_i(g-ij)\cap C^c))}
\end{equation}
and \emph{every} such $C$ satisfies $C\cap N_i(g-ij) \neq \emptyset$ and $i\notin C$ and $j\notin C$.
\end{claim}

\noindent{ \bf Proof of Claim \ref{c1}:}

We know from [S3] there exists $C \in \mathcal{C}(g-ij)$ such that
\begin{equation*}
\gamma<\frac{M(C)}{M(i\cup(N_i(g-ij)\cap C^c))}.
\end{equation*}

Suppose that some such $C$ has $C\cap N_i(g-ij)=\emptyset$.
This implies that $|C| > \gamma d_i$, and since $\gamma\geq 1$ and $d_i$ is maximal, this implies that $|C|=d_i+1$.
However, this is a contradiction since then all but one member of $C$ can defeat a remaining member (who necessarily has degree $d_i$ and thus only has connections
 to other members of $C$).  This follows since $d_i > \gamma$ given that $|C|=d_i +1> \gamma d_i$ and $d_i\geq 1$.

 Thus, any $C \in C\in\mathcal{C}(g-ij)$ satisfying (\ref{eqis3b}) must satisfy $C\cap N_i(g-ij)\neq \emptyset$.
 The fact that $i\notin C$ is by definition, and that $j\notin C$ is that otherwise we would violate [S1] (as $C$ would defeat $i$ in the network $g$ with $ij$ present).\eproof

\begin{claim}
\label{c2}
Consider $i$ of maximum degree and some $ij\in g$.
Consider any $C \in \mathcal{C}(g-ij)$ such that
\begin{equation*}
\gamma<\frac{M(C)}{M(i\cup(N_i(g-ij)\cap C^c))}.
\end{equation*}
It follows that
$$C \subset N_i(g) $$
and that\footnote{Note that this implies that $\frac{\gamma d_i}{1+\gamma}$ cannot be an integer.}
\begin{equation*}
|C | = \lceil \frac{\gamma d_i}{1+\gamma}\rceil \ \ {\rm and} \ \ |C | > \frac{\gamma d_i}{1+\gamma}.
\end{equation*}
Moreover, for \emph{any} $C\in \mathcal{C}(g)$ with $i\notin C$,
\begin{equation*}
|(C \cap N_i(g)) | \leq \lceil \frac{\gamma d_i}{1+\gamma}\rceil.
\end{equation*}
\end{claim}

\noindent{ \bf Proof of Claim \ref{c2}:}

Let $x=|C \cap N_i(g) |$, and let $y=|C \cap N_i(g)^c |$ be the number of members of $C$ who are not connected to $i$.  Then $\gamma<\frac{M(C)}{M(i\cup(N_i(g-ij)\cap C^c))}$ implies that
\begin{equation}
\label{eqz}
x+y >\gamma (1+d_i-1-x)=\gamma(d_i-x).
\end{equation}
Let $k\in C \cap N_i(g)$ (by Claim \ref{c1}).  [S1] implies that the remaining members of $C$ cannot defeat $k$ and so:
$$x+y -1 \leq \gamma (d_k+1 -(x+y-1)).$$
The fact that $d_k\leq d_i$ ($i$ is of maximal degree) and the two above inequalities imply that
$$\gamma (d_i-x)-1 < \gamma (d_i+2 -x-y),$$
or
$\gamma (y-2)< 1$.  Given that $\gamma\geq 1$ and $y$ is an integer, $\gamma (y-2)< 1$ implies that $y\leq 2$.   Now, let us argue that $y=0$.
Suppose to the contrary that $y=2$ (a similar argument will show that $y\ne 1$).  Let $k'$ and $k''$ be the countries in $C \cap N_i(g)^c$.   Consider the network $g+ik'$, and the clique
of $C'= (C\setminus \{k''\}) \cup \{i\}$, and note that $|C'|=|C|$.
By (\ref{eqz}) with $y=2$ and $x=|C|-2$, we know that $|C| > \gamma (d_i - (|C|-2))$.
But then, since $i$ has maximal degree and $|C'|=|C|$, it follows that  $|C'|> \gamma (d_{k''} -|C'| +2)$.  However, this contradicts [S2], since then $i$ and $k'$ can form a link
and the resulting clique $C'$ defeats $k''$. (To prove $y\ne 1$, take $k''$ to be any country in $C$ not equal to $k'$.)

Next, using (\ref{eqz}) and $y=0$ it then follows that
$$x(1+\gamma) >  \gamma d_i$$
or
$$x > \frac{\gamma d_i}{1+\gamma}.$$
Given that $\gamma\geq 1$ and $x$ is an integer, this implies that
\begin{equation}
\label{eqx}
x=|C | \geq \lceil \frac{\gamma d_i}{1+\gamma}\rceil  .
\end{equation}

To see the last part of the claim, let $z=|(C \cap N_i(g))\setminus \{i\} |$.
By [S1] (with $C$ not defeating $i$):
$$z\leq |C| \leq \gamma (d_i+1-z)$$
and so
$$z\leq \frac{\gamma (d_i+1)}{1+\gamma},$$
which, given that $z$ is an integer, implies that
\begin{equation*}
z=|(C \cap N_i(g))\setminus \{i\} | \leq \lceil \frac{\gamma d_i}{1+\gamma}\rceil.
\end{equation*}
as claimed.

The second part of the claim then follows from the last part of the claim and (\ref{eqx}).\eproof

\begin{claim}
\label{c3}
Consider $i$ of maximal degree and $ij\in g$.
It must be that
$$N_i(g)\setminus \{j\} \neq N_j(g)\setminus \{i\}.$$
\end{claim}

\noindent{ \bf Proof of Claim \ref{c3}:}

Consider $i$ of maximum degree and some $ij\in g$.
By Claim \ref{c1}, there exists $C \in \mathcal{C}(g-ij)$ such that
\begin{equation}
\label{eqis3b}
\gamma<\frac{M(C)}{M(i\cup(N_i(g-ij)\cap C^c))}
\end{equation}
and \emph{every} such $C$ satisfies $C\cap N_i(g-ij) \neq \emptyset$ and $i\notin C$ and $j\notin C$.

By Claim \ref{c2}, $C \subset N_i(g-ij)$.  If $N_i(g)\setminus \{j\} = N_j(g)\setminus \{i\}$,
 then $C \cup \{j\}$ is also a clique.  But then $|C \cup \{j\} |> |C| $, and we violate the last part of Claim
 \ref{c2}.\eproof



\begin{claim}
\label{c4}
There are no nonempty CN-war-stable networks (when $\gamma\geq 1$.
\end{claim}

\noindent{ \bf Proof of Claim \ref{c4}:}

Let $i$ be of maximum degree.
In order to satisfy [S3], it must be that for each $j\in N_i(g)$ there exists $C_j\in C(g)$ such that
\begin{equation*}
\gamma<\frac{M(C_j)}{M(i\cup(N_i(g-ij)\cap C^c_j))}.
\end{equation*}
Moreover, it follows from Claim
 \ref{c1} that each $C_j$ can be taken to lie in
$C(g-ij)$.
From Claim \ref{c2}, each such $C_j$ is such that $C_j\subset N_i(g)$ and $|C_j|= \lceil \frac{\gamma d_i}{1+\gamma}\rceil > \frac{d_i}{2}$.

Moreover, by Claim \ref{c2} it must be that for each $j'\in N_i(g)$, $\cup_{C_j\ni j'} C_j\ne N_i(g)$.
This follows since $i$ is of maximum degree and otherwise this would imply that
$N_{j'}(g)\setminus \{i\} = N_i\setminus \{j'\}$, contradicting Claim \ref{c3}.

Finding such sets $C_j$ for each $j$ in $N_i(g)$ thus becomes the following
combinatorics problem: create subsets $\{C_1,C_2,\ldots, C_S\}$ of a set $M=\{1,2,\ldots,d_i\}$ of $d_i$ elements (the neighbors of $i$) such that:
\begin{enumerate}
\item $\forall C_s$, $|C_s|=x>\frac{d}{2}$,

\item $\forall j\in M$, $\exists C_s$ such that $j\notin C_s$,

\item $\forall j\in M$, $\cup_{C_s\ni j} C_s\ne M$, and

\item $\not\exists D\subset M$ such that $\forall \{k,j\}\subset D$, $\exists C_s$ such that $\{k,j\}\subset C_s$ and $|D|>x$.

\end{enumerate}

4 follows from Claim \ref{c2} as otherwise $D$ would be a clique of size larger than $x=\lceil \frac{\gamma d_i}{1+\gamma}\rceil $.

We now show that such a collection of subsets is not possible. To do this, we start with just the set $C_1$ and see what implications hold as we consider each additional $C_s$, ultimately reaching a contradiction. For reference, we introduce the three new series of sets: $\{W_s\}_{s=1}^S$, $\{Y_s\}_{s=1}^S$, and $\{Z_s\}_{s=1}^S$. $W_s$ are the set of elements of $M$ which have been in at least one of the sets $C_1,\ldots,C_s$ (i.e. $W_s=\cup_{i=1}^s C_s$). $Y_s$ are the set of elements of $M$ which have been in all of the sets $C_1,\ldots,C_s$ (i.e. $Y_s=\cap_{i=1}^s C_s$). $Z_s$ are the set of elements of $M$ which have been in none of $C_1,\ldots,C_s$ (i.e. $Z_s=M\setminus W_s$).

Let us now complete the proof.
Note that, if a set of subsets $\{C_1,\ldots, C_S\}$ satisfying 1-4 existed, then $Y_S=\emptyset$ follows from point 2  since every element of $M$ has some $C_s$ that doesn't contain it. Note also that with each additional $C_s$, $W_s$ (weakly) grows larger while $Y_s$ and $Z_s$ (weakly) grow smaller.

To complete the proof we show that
$$|Y_{s-1}\setminus Y_s|\leq |Z_{s-1}\setminus Z_s|$$
and that
$$|Y_1|> |Z_1|.$$
Together these imply that $Y_S\neq \emptyset$, which is then a contradiction.

We start with $Y_1=W_1=C_1$.  Thus,  $|Y_1|=|W_1|=x>d_i-x=|Z_1|$ since $x>\frac{d_i}{2}$.
So, let us show that $|Y_{s-1}\setminus Y_s|\leq |Z_{s-1}\setminus Z_s|$.
At each subsequent addition of a $C_s$, either $C_s\cap Y_{s-1}=Y_{s-1}$ or $C_s\cap Y_{s-1}\subsetneqq Y_{s-1}$. In the first case, the result follows directly since then by definition $Y_{s+1}=Y_s$ and $0 \leq |Z_{s-1}\setminus Z_s|$.
So consider the second case.
In the second case, we show that $|Y_{s-1}| - |Y_{s-1}\cap C_s|\leq |Z_{s-1}\setminus Z_s|$.
Let $A= Y_{s-1}\setminus Y_s$ be the set of $j$ such that $j\in\cap_{i=1}^{s-1}C_i$ but $j\notin C_s$.
We show that $|C_s\cap Z_{s-1}|\geq |A|$ - that is, $C_s$ contains at least as many elements which aren't in any $C_{s'}$, $s'<s$ as there are elements of which are in every $C_{s'}$, $s'<s$ but not in $C_s$ (this establishes our result since $|C_s\cap Z_{s-1}|=|Z_{s-1}\setminus Z_s|$). To see this, suppose it weren't true. That is, suppose $|C_s\cap Z_{s-1}|< |A|$. Then, we would have set $D=(C_s\setminus Z_{s-1})\cup A$ of size at least $x+1$ that would contradict 4.
To see that the size is at least $x+1$, note that $C_s$ has by assumption $x$ members; by excluding $C_s$'s intersection with $Z_{s-1}$, we are excluding at most $|A|-1$ members of $C_s$, and adding in the $|A|$ elements of $A$. To see that $D$ satisfies the conditions of 4, note that any pair of of elements $k,\,j$ both of elements will satisfy $\{k,j\}\in C_1$. Likewise, any pair of elements $k,\,j$ both in $C_s\setminus Z_{s-1}$ will satisfy $\{k,j\}\in C_s$. Finally, any pair of elements $k,\,j$ with $k\in A$, $j\in (C_s\setminus Z_{s-1})\setminus A$ will satisfy $\{k,j\}\in C_{s'}$ for some $s'<s$ since $k$ is in all such $C_{s'}$ and since $j\in C_s\setminus Z_{s-1}\subset W_{s-1}$, $j$ is in at least one such $C_{s'}$. So, we have found a set of size at least $x+1$ satisfying the restrictions of point 4.  The contradiction establishes the impossibility of satisfying the combinatorics problem and thus the claim.\eproof

\medskip

We have thus shown that there exist no nonempty CN-war stable networks when $\gamma\geq 1$.
The final part of the theorem, that there are no CN-war stable networks if $\gamma<2$, then follows
from the fact that the empty network fails to satisfy [S2] if $\gamma< 2$ (but satisfies it if $\gamma \geq 2$),
as already established.\eproof

\bigskip

\bigskip
\noindent {\bf Proof of Theorem \ref{econstable}:}
We apply Proposition \ref{pairwisesuff}.
It is clear that any network that is $d^*$ regular is pairwise stable.
Thus, we need only show that no country is vulnerable despite trade and that this remains true with the addition of
any link.  For the first part of the proposition we also need to show that this is true regardless of $\delta E(\cdot)$ for at
least some $d^*$-regular networks.   For the second part of the proposition, we need to show this is true under the
given assumption on $ E(\cdot)$, but for any $d^*$ regular network.

First, note that no country $i$ is vulnerable to any coalition $C$ that does not include any of its neighbors (even if this comes from the addition of a link not involving any neighbors), since under either part of the theorem $\gamma \geq\frac{d^*+1}{d^*-1}> \frac{d^*+2}{d^*+1}$.  Thus, we need only verify
vulnerability to a coalition that involves at least one neighbor, and might possibly involve the addition of a link.

So, consider a country $i$ and a coalition $C$ that involves at least one of its neighbors.
Under the first part of the theorem, the maximum strength of the coalition (involving adding a link) would be $d^*+2$ (if the center is not one of $i$'s neighbors) and then the defending coalition would involve at least $d^*-k$ members, or else the center is one of $i$'s neighbors in which case the strength is at most $d^*+1$ and the defense involves at least $d^*-k-1$ members.  Given that $\gamma\geq
\frac{d^*+1}{d^*-k-1}$, it follows that $\gamma\geq
\frac{d^*+2}{d^*-k}$, and so $i$ is not vulnerable in either case.

Under the second part of the theorem, if any of those neighbors of $i$ in $C$ still has only $d^*$ links, then since $\frac{ E (d^*)}{2}  \leq  f(d^*) - f(d^*-1) - c$, and the attacking coalition must involve at least two countries (given $\gamma$ and $i$ at a minimum defending itself), then the country will not be willing to follow through with the attack of $i$ since it will lose a link.   Thus, all of $i$'s neighbors in the coalition $C$ must be gaining a link.  This implies that the coalition involves at most two of $i$'s neighbors, but then since $\gamma \geq\frac{d^*+1}{d^*-1}\geq \frac{d^*+2}{d^*}$, the attacking coalition
cannot defeat $i$ and its remaining neighbors, regardless of whether it involves one or two of $i$'s neighbors.\eproof

\section{Appendix:  Snapshots of Networks of Alliances: 1815 to 2000}\label{snapshots}

\cleardoublepage
\thispagestyle{empty}

\cleardoublepage
\thispagestyle{empty}

\cleardoublepage
\thispagestyle{empty}

\cleardoublepage
\thispagestyle{empty}


\cleardoublepage
\thispagestyle{empty}

\cleardoublepage
\thispagestyle{empty}

\cleardoublepage

\newpage

\bigskip

\setcounter{page}{1}

\section*{For Online Publication - Supplementary Material:  Stability Under CN-Vulnerability}
\label{CN-vuln}

\begin{figure}
	\centering
	\begin{minipage}[b]{0.5\textwidth}
		\centering
		\includegraphics[page=2, trim=0mm 90mm 0mm 100mm, scale=0.35]{WarNetworks_exampleGamma_lt_1.pdf}
		\caption{Stable Alliance Network}
		\label{figure:stable_net}
	\end{minipage}%
	\begin{minipage}[b]{0.5\textwidth}
		\centering
		\includegraphics[page=1, trim=0mm 90mm 0mm 100mm, scale=0.35]{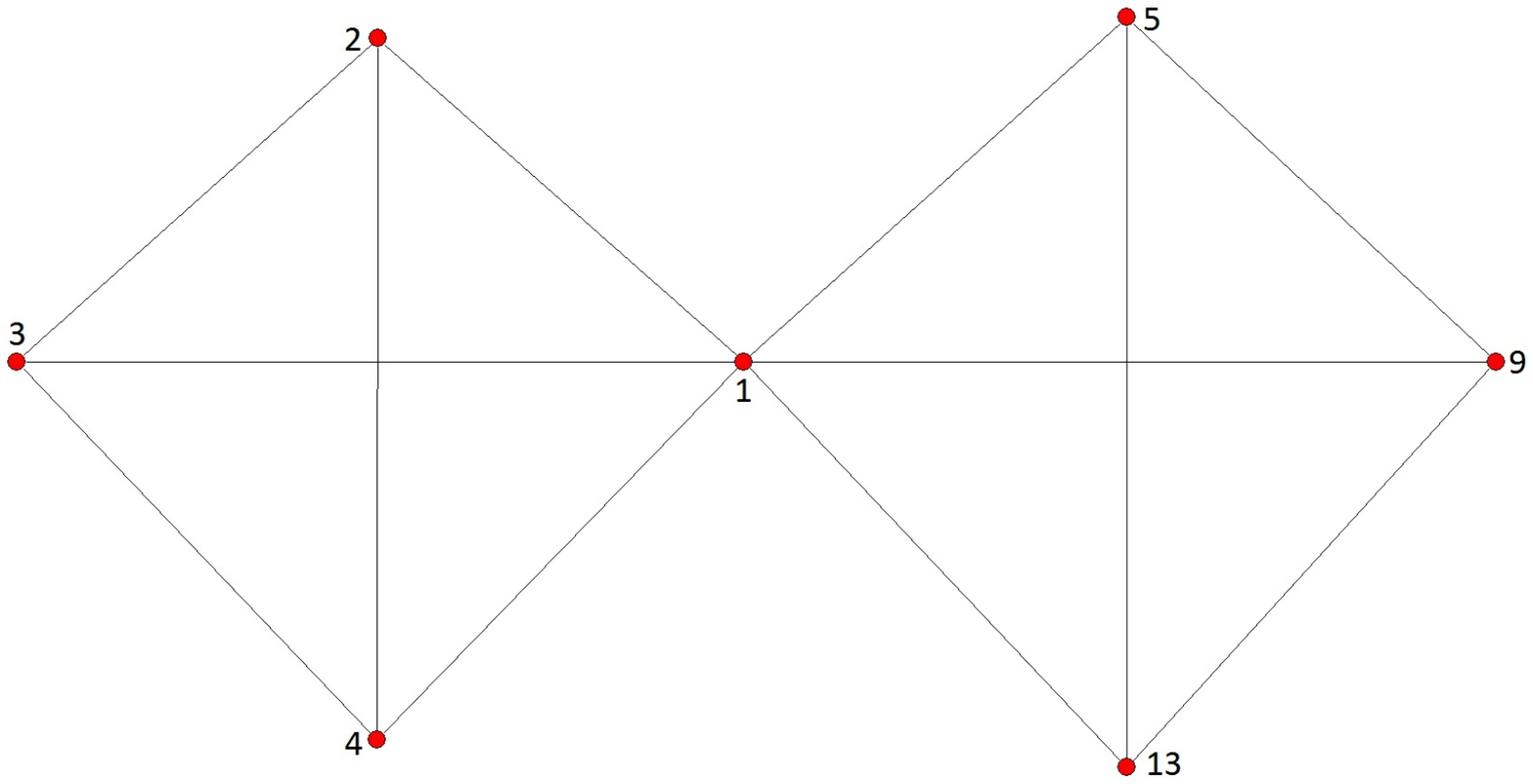}
		\caption{Close-up of Country 1 and Its Allies}
		\label{figure:stable_net_closeup}
	\end{minipage}
\end{figure}

To illustrate the point from Footnote \ref{footnote}, a war-stable configuration when $\gamma<1$ under CN-vulnerability is pictured in Figure \ref{figure:stable_net}.

Figure \ref{figure:stable_net_closeup} is a subgraph of Figure \ref{figure:stable_net} focusing on country 1, its neighbors, and the cliques it is involved in (note that each other country in \ref{figure:stable_net_closeup} has 3 other neighbors in the larger network, with each pair of countries one from the 1-2-3-4 clique and the other from the 1-5-9-13 clique sharing one neighbor in common).   The network is isomorphic, so that country 1 is perfectly representative of what is faced by all countries.  It can be easily verified that this network has no vulnerable countries if $\gamma\geq\frac{3}{4}$.  For instance if  country 1 were attacked by 2,3,4, it would be defended by 5,9,13 and so there would be 3 attackers and four defenders (counting 1) - and the defenders would win if $\gamma\geq \frac{3}{4}$.  Any other cliques outside of 1's neighbors would face 7 defenders if they attacked 1, and so would also lose.
It is easy to check that this network is stable against the addition of new links, since adding a new link does not increase the size of any clique, it just adds a new pair and pairs cannot win when attacking any country.   Thus, [S2] is easily checked.  So, it remains to check [S3].
If country 1 drops one of its links, e.g. with country 2, it will be vulnerable if $\gamma<1$ (country 1 could be attacked by countries 5, 9, and 13 and only defended by 3 and 4). Since similar arguments can be made for every other country (so that [S3] is satisfied for the network as a whole), if $\gamma\in[\frac{3}{4}, 1)$, the network is stable.

Similar examples can be constructed for even lower $\gamma$s, by having countries be part of more separate cliques.  For instance having each country be part of 3 separate cliques of size 4 (adding one more clique to the right hand side of Figure \ref{figure:stable_net} and for every country) would lead to a stable network for $\gamma \in [ \frac{3}{7},\frac{1}{2})$.  By varying the sizes of the cliques and the number of cliques that each country is in involved in, even arbitrarily small $\gamma$s nonempty stable networks can be found for large enough $n$.

\begin{proposition}
\label{gammalow}
Consider any $\gamma<1$ and the uniform strength case in which $M_i=\overline{M}\;\forall i$.  There exists a large enough $n$ such that there is a nonempty
war-stable network under CN-vulnerablity in which every country has at least one alliance.
\end{proposition}

Proposition \ref{gammalow} provides an interesting contrast to Theorem \ref{IS-Robust}.  With an offensive advantage, stable networks exist under CN-vulnerability as the offensive advantage provides incentives for countries to maintain relationships as a deterrent that they might sever in conditions where there is a defensive advantage.

\bigskip

\noindent {\bf Proof of Proposition \ref{gammalow}:}
We prove by constructing a network such that every country $i$ is a member of 2 cliques, each of size 4, and with $i$ being the only country in both cliques. Further, we let $i$ have $a$ additional neighbors, none of whom are connected to each other or to other countries in the cliques $i$ is in. The required number of countries will be $32a$.

Start with a network with 16 countries, each involved in 2 cliques of size 4. To construct it, take four copies of $k_4$ (the complete network on 4 vertices). In each copy of $k_4$, label the countries 1 through 4. Then, connect all countries labelled 1 to each other, all countries labelled 2 to each other, and so on.

To construct the final network $g$, take this network on 16 countries, relabel the countries ``a'' through ``p'' (the $16^{th}$ letter) arbitrarily. Then, create $2a$ copies of the network on 16 countries, numbering each copy 1 through $2a$. Label each vertex by (number of the position in the first network, letter of the position in the second network). There will be $32a$ labels, from (1,a) through ($32a$,p). Now connect each country to all other countries which agree on letter but differ on parity of the number (e.g. connect (1,a) with (3,a), (5,a), and so on; in essence, each subset of countries of the same letter forms a $a$-regular network with no cliques larger than 2).

In this final network, there are only cliques of size 4 (in the copies of the starting network) and of size 2. Further, adding any link to the network can create a new clique of size at most 3, involving no more than 2 of any given country's neighbors. Since each country can already be threatened by a clique of 3 of its neighbors, we may safely ignore what restrictions on $\gamma$ the requirement that no country be vulnerable in $g+ij$. For no country $i$ to be vulnerable in $g$, we need $\gamma\geq\frac{3}{4+a}$ ($i$ can be attacked by 3 of its neighbors connected in one of the size 4 cliques $i$ is in, and $i$ is defended by its neighbors in the other size 4 clique it is in, as well as its $a$ other neighbors). To prevent $i$ from wanting to drop a link, we need $\gamma<\frac{3}{3+a}$. Combining, we need $\gamma\in[\frac{3}{4+a},\frac{3}{3+a})$. For $a=0$, this interval is $[\frac{3}{4},1)$. Rearranging the interval, we need $\frac{1}{\gamma}\in(1+\frac{a}{3},\frac{4}{3}+\frac{a}{3}]$, so that for any $\gamma<1$, we can take the largest $a$ satisfying $a<\frac{3}{\gamma}-3$, satisfying the lower bound, and then the upper bound will necessarily be satisfied.\eproof

\subsection{Heterogeneous Military Strengths under CN-vulnerability}

Although we have not been able to find any examples of nonempty war-stable networks under CN-vulnerability when $\gamma\geq 1$ and conjecture that the result extends generally, it appears quite
difficult to prove.  The complexity of the proof of Theorem \ref{IS-Robust} shows the logic is involved.
Parts of that proof are combinatorial in nature, and not directly extendable to asymmetries.

Nonetheless, we can show that the CN-result in Theorem \ref{IS-Robust} is robust in the sense that it holds for any (unequal) military strengths in an open neighborhood around equal ones.   The question of whether the open neighborhood can be expanded to the full set remains open.

Let us say that $\gamma$ has the {\sl no-tie} property relative to $n$ if there does not exist any positive integers $m_1$ and $m_2$ for which $m_1+m_2\leq n$ and
$m_1=\gamma m_2$.   This is clearly a generic property, as it holds directly for all irrational levels of $\gamma$, as well as
many rational levels - and a full measure set of values.

\begin{theorem}
\label{ISprime}
Let $n\geq 3$ and consider $\gamma$ that satisfies the no-tie property relative to $n$ and some base military strength $\overline{M}$.
There exists an open set of military strengths $\mathcal{M}$ with $\overline{M}\in \mathcal{M}$ for which the following holds. If $1\leq \gamma <2$
and $(M_1,\ldots, M_n)\in \mathcal{M}^n$, then there are no CN-war-stable networks.
If $\gamma\geq 2$ and $(M_1,\ldots, M_n)\in \mathcal{M}^n$, then the unique CN-war-stable network is the empty network.
\end{theorem}

Theorem \ref{ISprime} shows that Theorem \ref{IS-Robust} is robust to some heterogeneity in military strengths.

\bigskip

\noindent{\bf Proof of Theorem \ref{ISprime}:}
The proof follows easily from the proof of Theorem \ref{IS-Robust} simply noting that if $\gamma$ admits no ties, then there is an open set of
military strengths for $M(C)>\gamma M(C')$ for any coalitions when all countries have strengths within $\mathcal{M}$ if and only if the same is true when all have strength $\overline{M}$.\eproof

\subsection{CN-War and Trade Stability}

In illustrating Footnote \ref{econstable1}, an example illustrating the existence of CN-war-and-trade stability appears in Figure \ref{figure:stable_net} in which $d^*=6$, and which is in fact strongly CN-war and trade stable for all $\gamma\geq \frac{3}{4}$.

The quilts take advantage of the fact that no clique is large enough to outweigh any country and its remaining allies, and yet the alliances are held in place by trading concerns so that no alliances are deleted.
Thus, the quilts leverage the economic incentives to maintain links, thereby overcoming the sparsity issues that can prevent stability.

\begin{figure}
	\centering
	\includegraphics[trim = 0mm 90mm 0 100mm, scale=0.4]{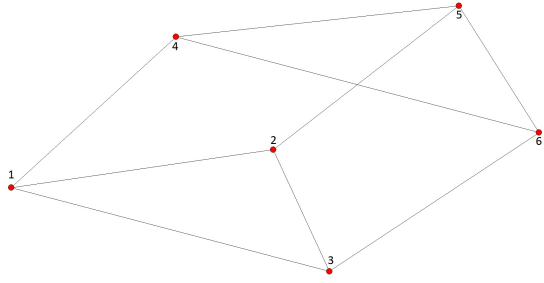}
	\caption{An alliance network}
	\label{figure:example_connected_triangles}
\end{figure}

Another aspect that trade incentives bring is that they reduce incentives for members of a clique to follow through with attacks.
For example, there are networks that are war and trade stable, but for which there is a $ij\notin g$ such that some country is vulnerable despite trade at $g+ij$. For instance, let $d^*=3$ and the number of countries be 6. Let $g$ be formed of two triangles, with each country in one triangle connected to exactly one country in the other triangle.
In particular, consider network $g=\{12,13,23,45,46,56, 14,25,36\}$, as pictured in Figure \ref{figure:example_connected_triangles}.
Let $\gamma<\frac{3}{2}$. Then, adding another link between the two triangles would make some country vulnerable despite trade at $g+ij$ (with an attacking coalition consisting of two countries from one triangle and one country from the other attacking the third country from the first triangle).
In particular, if $15$ is added, then $125$ can defeat 3 who only has ally 6 remaining.
However, there are several possible deterrents to this attack.  One is that $\delta$ is low enough so that the gains from conquering 3 are not worth it for 2 who then only has 2 trading partners in the resulting network.  Similarly it might be that $E_3$ is small.
Another possibility is that 5 might not wish to have four allies instead of three, and that the resulting gains from conquering country 3 are not worth the cost of an additional ally.

\end{document}